\documentclass[amsmath,amssymb,superscriptaddress]{revtex4}
\usepackage{graphicx}% Include figure files
\usepackage{dcolumn}% Align table columns on decimal point

\def\be{\begin{equation}}
\def\ee{\end{equation}}
\def\beq{\begin{eqnarray}}
\def\eeq{\end{eqnarray}}
\def\n{\nonumber}
\usepackage{bm}% bold math
\usepackage{graphicx}
\usepackage{dcolumn}
\usepackage{amsmath}
\usepackage[latin1]{inputenc}
\usepackage{graphicx, psfrag}
\usepackage{amssymb}
\usepackage[colorlinks=true, citecolor=blue, urlcolor = blue, linkcolor= red, bookmarks=true]{hyperref}
\usepackage{float}
\usepackage{amsmath}
\usepackage{amsfonts}
\usepackage{dcolumn}
\usepackage{hyperref}
\usepackage{subfigure}
\usepackage{pgfplots}
\usepackage{epstopdf}
\usepackage{booktabs}
\DeclareMathOperator{\sech}{sech}
\begin{document}

% Use the \preprint command to place your local institutional report
% number in the upper righthand corner of the title page in preprint mode.
% Multiple \preprint commands are allowed.
% Use the 'preprintnumbers' class option to override journal defaults
% to display numbers if necessary
%\preprint{}

%Title of paper
\title{All Conformally Flat Einstein--Gauss--Bonnet static  Metrics}

% repeat the \author .. \affiliation  etc. as needed
% \email, \thanks, \homepage, \altaffiliation all apply to the current
% author. Explanatory text should go in the []'s, actual e-mail
% address or url should go in the {}'s for \email and \homepage.
% Please use the appropriate macro for each each type of information

% \affiliation command applies to all authors since the last
% \affiliation command. The \affiliation command should follow the
% other information
% \affiliation can be followed by \email, \homepage, \thanks as well.
\author{Sudan Hansraj} \email[]{hansrajs@ukzn.ac.za}
\author{Megandhren Govender} \email[]{megandhreng@dut.ac.za}
\author{Ayan Banerjee} \email[]{ayanbanerjeemath@gmail.com}
\author{Njabulo Mkhize} \email[]{mkhizen18@gmail.com} 

\affiliation{Astrophysics and Cosmology Research Unit, School of Mathematics, Statistics and Computer Science, University of KwaZulu--Natal, Private Bag X54001, Durban 4000, South Africa}

%\homepage[]{Your web page}
%\thanks{}
%Collaboration name if desired (requires use of superscriptaddress
%option in \documentclass). \noaffiliation is required (may also be
%used with the \author command).
%\collaboration can be followed by \email, \homepage, \thanks as well.
%\collaboration{}
%\noaffiliation

\date{\today}

\begin{abstract}
% insert abstract here

It is known  that the standard  Schwarzschild interior metric is conformally flat and generates a constant density sphere in any spacetime dimension in Einstein and Einstein--Gauss--Bonnet gravity. This motivates the questions: In EGB does the conformal flatness criterion yield the Schwarzschild metric? Does the assumption of constant density generate the Schwarzschild interior spacetime?  The answer to both questions turn out in the negative in general. In the case of the constant density sphere, a generalised Schwarzschild metric emerges.  When we invoke the conformal flatness condition  the Schwarschild interior solution is obtained as one solution and another metric which does not yield a constant density hypersphere in EGB theory is found. For the latter solution one of the gravitational metrics is obtained explicitly while the other is determined up to quadratures in 5 and 6 dimensions. The physical properties of these new solutions are studied  with the use of numerical methods and a parameter space is located for which both models display pleasing physical behaviour.

\end{abstract}

% insert suggested PACS numbers in braces on next line
% insert suggested keywords - APS authors don't need to do this
%\keywords{}

%\maketitle must follow title, authors, abstract, \pacs, and \keywords
\maketitle

% body of paper here - Use proper section commands
% References should be done using the \cite, \ref, and \label commands

\section{Introduction}

% Put \label in argument of \section for cross-referencing
%\section{\label{}}
%\subsection{}
%\subsubsection{}

In the standard theory of gravity due to Einstein all conformally flat spacetimes are known.  They are either the Schwarzcshild interior metric in the case of no expansion or the Stephani universes when expansion is permitted \cite{steph1,steph2,hans1}. A similar  result is not known in the more complicated Einstein-Gauss--Bonnet (EGB)  theory so this is one of the motivations behind this work. Conformal structures are important  in gravitational field theory as conformal symmetries generate constants of the motion or conserved quantities along null geodesics. Conformal flatness is characterised by the vanishing of the Weyl tensor and physically this means that a spacetime is conformal to the Minkowski metric at spatial infinity. This is a reasonable restriction when it comes to modelling relativistic compact objects such as neutron stars, white dwarfs or cold planets. It is already known both in the four dimensional general theory of relativity as well as its extensions to higher dimensions, that that  the interior Schwarzschild metric is both a necessary and sufficient  condition for conformal flatness \cite{hansrajjmp}. Whether this holds when higher curvature effects are at play will be discussed herein.  

Recently it has been demonstrated by Dadhich {\it{et al}} \cite{dad-mol-khug} that the Schwarzschild interior metric is universal as a constant density solution in Lovelock gravity of which the EGB theory is a special case. These authors claimed necessity and sufficiency.  The assumption of the Schwarzschild interior spacetime does indeed generate a constant density fluid. However, the converse is not true in general. The assumption of constant density generates a solution which generalises the Schwarzschild interior spacetime. That is, the prescription of  constant density is only a necessary but not sufficient condition for the Schwarzschild interior metric.   The aforesaid authors argued that an integration constant must vanish in order to ensure regularity at the stellar centre and consequently the solution reduces to the Schwarzschild metric.  We shall show that this is not required in the five dimensional constant density case which possesses no central  singularity however the six dimensional incompressible hypersphere metric does have a persistent singularity at the stellar centre.  Nevertheless, in neither case is there a basis to remove integration constants arbitrarily. These constants are necessary and may be determined through the boundary conditions.
 These aspects are being mentioned in the context of conformal flatness because the Schwarzschild solution is conformally flat but it needs to be checked whether all conformally flat static metrics are necessarily the Schwarzschild interior solution.  It must be remarked that ordinarily it is not advisable to set integration constants to zero arbitrarily since these must be settled by matching of the interior and exterior spacetime across a common hypersurface. Effectively deciding the value of constants of integration rearly in modelling  runs the risk of over-determining the model.  This option may only succeed when there is no boundary and a universe model is in evidence. This is the case for the isothermal spherically symmetric universe of \cite{saslaw} which was clearly over-determined but successful. In this same spirit, Hansraj and Moodly \cite{hans-moodly} showed that the dual requirement of conformal flatness and spacetime being of embedding class one, gives a null-result. 

At the outset, we sketch motivations for considering higher dimensional spacetimes and the importance of the EGB theory in gravitational field theory. 
Interest in higher dimensional spacetimes originated with the seminal works of Kaluza \cite{kaluza} and Klein \cite{klein}. A five dimensional manifold was introduced in the context of the Einstein--Maxwell equations and now there were 15 nontrivial components as opposed to 10 in the standard four dimensions. Four of the  components of the metric tensor were associated with the electromagnetic field while ten were connected to the usual 4 dimensional space. The remaining component was given the interpretation  of a scalar field termed a scalaron or dilatonic field. Subsequently many other ideas emerged that considered higher dimensions. In fact, in the quantum world,  it is now understood that the most promising idea superstring theory and its generalisation $M$-theory both require dimensions of the order of 10 or 11 at least. Additionally, brane-world cosmology \cite{maartens}  requires spacetimes of dimension five. But what would be the explanation for our inability to access extra spatial dimensions? These are believed to be topologically curled into microscopic circles of very small size yet they exert an influence on the gravitational field  \cite{kaluza,klein}. Probing gravitational waves for information on extra dimensions was considered in a recent paper \cite{yu} that reported the non-detection of large extra dimensions. Earlier theories predicting large extra dimensions were tested through the Large Hadron Collider (LHC) experiment but found to be inconclusive \cite{lhc1,lhc2,lhc3,lhc4}.    This however, does not eliminate the possibility of small extra dimensions \cite{mack} presumed to be of the planck length.  It must be remarked that the extra dimensions are actually angular dimensions so they do not necessarily manifest overtly.  In light of these, investigations into higher dimensional spaces are justifiable.

Why is the EGB gravity proposal of immense importance? Firstly it offers a natural generalization of Einstein's theory to higher dimensions with the inclusion of higher curvature terms but without violating well established ingredients of gravitational field theory namely the Bianchi identities, diffeomorphism invariance and second order (ghost--free) equations of motion. The energy conservation conditions also hold in the usual way adjusted for extra dimensions.  In fact EGB is superseded by its generalisation Lovelock \cite{lov1,lov2} theory which has the same aforementioned properties. The $Nth$ order Lovelock Lagrangian is constructed from invariants polynomial in the Riemann tensor, Ricci tensor and the Ricci scalar and to second order ($N = 2$) the EGB special case consists of quadratic invariants.  The critical spacetime dimensions $d$ in Lovelock theory are $d = 2N + 1$ and $d = 2N + 2$. In particular it is sufficient to consider only dimensions 5 and 6 in studying EGB spacetimes \cite{dad-ghosh-jhingan}. It is also well known that the Gauss-Bonnet term features in heterotic string theory and the coupling constant carries the meaning of the string tension in that area. This is also a good reason to probe the behaviour of higher curvature invariants in gravitational field theory in view of the long time project to merge quantum field theory, of which string theory is a leading candidate, and gravitational physics. 

The exterior solution for EGB gravity was found by Boulware and Deser \cite{boul} in 1985 and a year later Wiltshire promoted the solution to include the effects of the electrostatic field \cite{wilt}.  The constant density solution of Dadhich {\it{et al}} \cite{dad-mol-khug} was the first interior solution reported and happened to be the Schwarzschild solution independent of the dimension. As in Einstein theory the solution suffers physical pathologies such as an infinite speed of sound. The square of the speed of sound is calculated with the formula $\frac{dp}{d\rho}$ where $p$ and $\rho$ are the pressure and density respectively. A thorough discussion of the relativistic sound speed was considered by Ellis {\it {et al}} \cite{ellis1}.  Clearly a constant density $\rho$ renders the sound speed meaningless. Kang {\it{et al}} \cite{kang} proposed a solution however, the metric demanded a further integration to fully reveal the nature of the spacetime manifold. The first reported explicit exact solutions that satisfied elementary conditions for physical plausibility were generated in \cite{hans-maha,maha-chil,chilambwe-hansraj}. It is seriously difficult to locate exact solutions for perfect fluid matter in EGB because the extra curvature terms make the governing differential equations intractable.  An additional solution for constant potentials in six dimensional EGB spacetimes was found in \cite{hans-maha-chil} and recently Hansraj and Mkhize generated a physically viable six dimensional model with variable potentials and density \cite{hans-mkhize}. A greater number  of the extra curvature terms survive in 6D as opposed to 5D making the differential equations even more difficult to work with. However, the effects of extra curvature may be studied more efficiently compared to the 5D case where such terms are suppressed.   

Employing an equation of state (EoS) to describe static compact objects in classical general relativity and modified gravity theories have proved not to be fruitful. The imposition of a barotropic equation of state of the form $p = p(\rho)$ reduces the problem of finding exact solutions to the Einstein field or the EGB equations to a single-generating function when pressure isotropy is dropped. Requiring the vanishing of the Weyl stresses at each interior point of the fluid sphere completes the gravitational description of the model. These models have been shown to describe compact objects such as pulsars, neutron stars and colour-flavoured locked-in quark stars \cite{weyl1,weyl2,weyl3,Banerjee:2020stc,Singh:2020bdv}. In fact, some models predicting the existence of self-gravitating anisotropic fluid spheres have been investigated in the literature. For example, see  \cite{Singh:2020cnu,Maurya:2020gjw,Rahaman:2020dgv,Singh:2020ebx,Tello-Ortiz:2019gcl} and the references therein.

The role of the Weyl tensor in dissipative collapse has been thoroughly investigated. Herrera and co-workers demonstrated that departure from hydrostatic equilibrium (or quasi-static equilibrium) is sensitive to changes in the conditions surrounding the Weyl tensor. In particular, in the case of conformally flat self-gravitating bodies, departure from equilibrium depends on the deviation from conformal flatness. It is well known that density inhomogeneities and pressure anisotropy influence the outcome of dissipative collapse. It has been shown that the thermal behaviour of radiating stars can be drastically altered in the presence of anisotropic stresses within the collapsing core. Furthermore, specific combinations of the Weyl tensor components, pressure anisotropy and dissipative fluxes (for example radial heat flux) give rise to density inhomogeneities within the collapsing fluid. The stability of the shear-free condition has come under scrutiny as it was shown that an initially shear-free configuration would evolve into a shearing-like regime due to a combination of density inhomogeneity, pressure anisotropy and dissipation\cite{prisco1}. Several exact models describing conformally flat radiating stars have been obtained. These solutions have been shown to obey the physical requirements necessary for dissipative collapse. Gravitational collapse from an initial static configuration described by the interior Schwarzschild solution has rewarded researchers richly in terms of insights into horizon formation, stability in the Newtonian and Post-Newtonian regimes and thermal evolution within the framework of extended irreversible thermodynamics\cite{stab1,stab2,stab3}. 

This paper is arranged as follows: After a brief sketch of the main ingredients of EGB theory in section \ref{sec2}, we derive the five dimensional field EGB field equations and consider the exterior metric in section \ref{sec3}. The incompressible hypersphere is examined in section \ref{sec4} and the new exact solution for conformal flatness together with its physical properties is presented in section \ref{sec5}.  In section \ref{sec6} we find he conformally flat metrics in six dimensions and study its main features. Finally in section \ref{sec7} we conclude by reiterating our main results.

\section{Einstein--Gauss--Bonnet Gravity}\label{sec2}

The Gauss--Bonnet action  is written as
\begin{equation}
S = \int \sqrt{-g} \left[ \frac{1}{2} \left(R - 2\Lambda + \alpha L_{GB}\right)\right] d^N x + S_{\mbox{ matter}}, \label{1}
\end{equation}
where $ \alpha $ is the Gauss--Bonnet coupling constant and $N$ is the spacetime dimension. The advantage  of the action $ L_{G B} $ is that despite the Lagrangian being quadratic in the Ricci tensor, Ricci scalar  and the Riemann tensor, the gravitational field equations are  second order quasilinear which is expected of a viable theory of gravity. The Gauss--Bonnet term exerts no influence for $ n \leq 4 $ but becomes dynamic for $ n > 4 $.

The EGB field equations may be written as
\begin{equation}
G_{a b} + \alpha H_{a b} = T_{a b},  \label{2}
\end{equation}
with metric signature $ (- + + + +) $ where $ G_{ab} $ is the Einstein tensor. The Lanczos tensor is given by
\begin{equation}
H_{a b} = 2 \left(R R_{a b} - 2 R_{a c}R^{c}_{\,\,b} - 2 R^{c d} R_{a c b d} + R^{c d e}_{\,\,\,\,\,\,\,\,a} R_{b c d e} \right) - \frac{1}{2} g_{a b} L_{G B},  \label{3}
\end{equation}
where the Lovelock term has the form

\begin{equation}
L_{G B} = R^2 - 4R_{a b} R^{a b} + R_{a b c d} R^{a b c d} .   \label{4}
\end{equation}
In what follows it will  be desirable to divide the analysis into the 5 and 6 dimensional cases separately as these embody different physical characteristics. The 5 dimensional case has the feature of eliminating some terms in the field equations which are dynamic only in 6-$d$ and consequently affect the physics of the distribution. 

\section{Field equations in 5-$d$}\label{sec3}

The generic 5--dimensional  line element for static spherically symmetric spacetimes is customarily written as 
\begin{equation}
ds^{2} = -e^{2 \nu} dt^{2} + e^{2 \lambda} dr^{2} + r^{2} \left( d\theta^{2} + \sin^{2} \theta d \phi^2 + \sin^{2} \theta \sin^{2} \phi d\psi^2 \right), \label{5}
\end{equation}
where $ \nu(r) $ and $ \lambda(r) $ are  the gravitational potentials. We utilise a comoving fluid velocity of the form $ u^a = e^{-\nu} \delta_{0}^{a} $ and the matter field is that of a perfect fluid with energy momentum tensor $ T_{a b} = (\rho + p) u_a u_b + p g_{a b} $. Accordingly the EGB field equations (\ref{2}) assume the form 
\begin{eqnarray}
\rho &=&- \frac{3}{e^{4 \lambda} r^{3}} \left( 4 \alpha \lambda ^{\prime} +  r e^{2 \lambda} -  r e^{4 \lambda} -  r^{2} e^{2 \lambda} \lambda ^{\prime} - 4 \alpha e^{2 \lambda} \lambda ^{\prime} \right),  \label{6a} \\ \nonumber \\
p_r &=&  \frac{3}{e^{4 \lambda} r^{3}} \left(-  r e^{4 \lambda} + \left( r^{2} \nu^{\prime} +  r + 4 \alpha \nu^{\prime} \right) e^{2 \lambda} - 3 \alpha \nu^{\prime} \right),  \label{6b} \\ \nonumber \\
p_T &=& \frac{1}{e^{4 \lambda} r^{2}} \left( -e^{4 \lambda} - 4 \alpha \nu^{\prime \prime} + 12 \alpha \nu^{\prime} \lambda^{\prime} - 4 \alpha \left( \nu^{\prime} \right)^{2}  \right) \nonumber \\
                 & \quad & + \frac{1}{e^{2 \lambda} r^{2}} \left(  1 - r^{2} \nu^{\prime} \lambda^{\prime} + 2 r \nu^{\prime} - 2 r \lambda^{\prime} + r^{2} \left( \nu^{\prime} \right)^{2}  \right) \nonumber \\
                 & \quad & + \frac{1}{e^{2 \lambda} r^{2}} \left(  r^{2} \nu^{\prime \prime} - 4 \alpha \nu^{\prime} \lambda^{\prime} + 4 \alpha \left( \nu^{\prime} \right) ^{2} + 4 \alpha \nu^{\prime \prime}   \right). \label{6c}
\end{eqnarray}
where $p_r$ and $p_T$ are the radial and tangential pressure respectively. Note that the system (\ref{6a})--(\ref{6c}) comprises three field equations in four unknowns for isotropic particle pressure $p_r = p_T$. This is similar to the standard Einstein case for spherically symmetric perfect fluids. In order to close the system of equations, it is necessary to stipulate one more condition. Traditionally, one of the metric potentials is prescribed in the hope of finding the remaining potential by integrating the equation of pressure isotropy. Alternatively mathematical insights such as shown by Tolman \cite{tolman} in rearranging the pressure isotropy condition in a convenient way and setting individual terms to vanish yielded eight classes of exact solutions five of which were new. The attempts at implementing physically reasonable assumptions such  as an equation of state has not worked in Einstein gravity and the works of Nilsson and Uggla \cite{nilsson1,nilsson2} in this direction had to be completed with numerical methods. Therefore it is not expected that an {\it{a priori}} equation of state $p = p(\rho)$ will succeed in the more complicated situation of EGB at hand.  Additionally, conditions such as being able to embed a spacetime in a higher dimensional geometry such as through use of the Kamarkar condition on the Riemann tensor components may also be an interesting route to follow \cite{Tello-Ortiz:2019gcl,Maurya:2017non,Ashraf:2020yyo,Mustafa:2020ikt}. Historically,  Schlaefli \cite{schlaefli} first raised the question of embedding a Riemannian space into a higher dimensional Euclidean space and Kamarkar \cite{kamarkar} established a condition on the components of the Riemann tensor. The Kamarkar condition was corrected in time to rule out the vanishing of one of the components \cite{pandey}.   The existence of symmetries such as conformal Killing vectors also offer an avenue to pursue. In our case we have elected to impose the condition for conformal flatness.    The  vacuum metric describing the gravitational field exterior to the 5--dimensional static perfect fluid may be described by the Boulware--Deser \cite{boul} spacetime as
\begin{equation}
ds^2 = - F(r) dt^2 + \frac{dr^2}{F(r)} + r^{2} \left( d\theta^{2} + \sin^{2} \theta d \phi + \sin^{2} \theta \sin^{2} \phi d\psi \right), \label{7}
\end{equation}
where
\be
F(r) = 1 + \frac{r^2}{4\alpha} \left( 1 - \sqrt{1 + \frac{8M\alpha}{r^4}} \right),
\ee
and where we have taken a negative sign before the square root since expansion in powers of $\alpha$ give the Schwarzschild solution when $\alpha $ approaches zero.  
In the above $ M $ is associated with the gravitational mass of the hypersphere. The exterior solution is  unique up to branch cuts.  

We invoke the transformation $ e^{2 \nu} = y^{2}(x) $, $ e^{-2 \lambda} = Z(x)  $ and $ x = C r^{2} $ ($ C $ being an arbitrary constant) which was utilised successfully by Durgapal and Bannerji \cite{dur-ban}, Finch and Skea \cite{finch-skea} and Hansraj and Maharaj \cite{hansraj-maharaj} to generate new exact solutions for neutral and charged isotropic spheres. The motivation for using this transformation lies in the fact that the equation of pressure isotropy may be expressed as a second order linear differential equations thus increasing our chances of locating exact solutions.    The field equations (\ref{6a})--(\ref{6c}) may be recast as
\begin{eqnarray}
3  \dot{Z} + \frac{3  (Z - 1) ( 1 - 4 \alpha C \dot{Z} )}{x} &=& \frac{\rho}{C}, \label{8a} \\ \nonumber \\
\frac{3  (Z - 1)}{x} + \frac{6  Z \dot{y}}{y} - \frac{24 \alpha C (Z - 1) Z \dot{y}}{x y} &=& \frac{p}{C}, \label{8b} \\ \nonumber \\
 2 x Z \left( 4 \alpha C [Z - 1] - x \right) \ddot{y}  - \left( x^{2} \dot{Z} + 4 \alpha C \left[ x \dot{Z} - 2 Z + 2 Z^{2} - 3 x Z \dot{Z} \right] \right) \dot{y} \nonumber \\ - \left( 1 + x \dot{Z} - Z \right) y &=& 0,  \label{8c}
\end{eqnarray}
where the last equation is  the equation of pressure isotropy. Equation (\ref{8c}) has been arranged as a second order differential equation in $ y $, which for some analyses in the 4--dimensional Einstein models, proves to be a useful form. Functional forms for $ Z(x) $ may be selected {\it a priori} so as to allow for the complete integration of the field equations. 
For the present work it should be noted that (\ref{8c}) may also be regarded as a first order ordinary differential equation in $ Z $, and may be expressed in the form
\begin{eqnarray}
\left( x^{2} \dot{y} + x y + 4 \alpha C x \dot{y} - 12 \alpha C x \dot{y} Z \right) \dot{Z} + 8 \alpha C \left( \dot{y} - x \ddot{y} \right) Z^{2} + \left( 2 x^{2} \ddot{y} + 8 \alpha C x \ddot{y} - 8 \alpha C \dot{y} - y \right) Z + y = 0. \label{9}
\end{eqnarray}
This is an Abel equation of the second kind for which few solutions are known. However note that  choosing forms for $ y $ should in theory result in the expressions for $ Z $ by integration. Therefore we seek choices for the metric potential $ y $ which will allow for a complete resolution of the geometrical and dynamical variables.

\section{Incompressible 5-$d$ hypersphere}\label{sec4}

Since it is true in standard Einstein gravity that conformal flatness, the interior Schwarzschild metric and constant energy density are all equivalent \cite{hansrajjmp}, it is important to examine the incompressible sphere in the present context.  The form $ Z=1+x $ which is the  Schwarzschild potential generates the  constant density distribution in higher dimensional Einstein gravity and also in higher curvature EGB gravity.  That is the Schwarzschild solution is a sufficient  condition for constant density.  This corroborates the result of Dadhich {\it{et al}} \cite{dad-mol-khug}  independent of spacetime dimension. However the question of necessaity arises. Does the assumption of constant density give the Schwarzschild metric? We shall analyse this problem below.  Moreover, it is known that the Schwarzschild interior metric is conformally flat. This is a purely geometric consequence irrespective of the gravity theory under consideration. In the Einstein theory it is already well known that invoking the conformal flatness condition generates the Schwarzschild metric.  It is therefore  natural to ask whether the prescription of conformal flatness to close the system of equations also   generates the interior Schwarzschild potential in EGB theory? Additionally does the general conformally flat EGB solution produce a constant density fluid. We examine these questions  separately for 5 and 6 dimensions.

We note that equation (\ref{8a}) can be integrated when the density is assumed to be constant $\rho = \rho_0$ and gives 
\begin{eqnarray}
Z(x) &=& 1 + \frac{x}{4\alpha C} \left( 1 \pm \sqrt{(3 - 4\alpha \rho_0) + \frac{c_1 \alpha^2 C^2}{x^2}} \right), \label{101}
\end{eqnarray}
where $c_1$ is an integration constant. Setting the constant density to vanish $\rho_0 = 0$ regains the vacuum solution (\ref{7}) of Boulware and Deser \cite{boul}.  In the event that $c_1 = 0$ the solution of Dadhich {\it{et al}} is regained and it is then demanded that the coupling constant satisfies $\alpha \leq \frac{3}{4\rho_0}$. In this case the form $Z = 1 + x$ arises but this is not in general true for EGB theory in general.  Substituting (\ref{101}) in the conformal flatness condition, to be displayed later, regretfully leads to an intractable differential which cannot be solved for now. So the $y$ potential remains unknown in this general case.  Note that Dadhich {\it{et al}} claim the integration constant should vanish for regularity at the centre however this is not a valid issue. If (\ref{101}) is written as 
\be Z = 1 + \frac{x}{4\alpha C}  \pm \frac{1}{4} \sqrt{  c_1 + \frac{x^2(3 - 4\alpha \rho_0)}{\alpha^2 C^2}}    
\ee  
then it is clear that  there is no singularity at the stellar centre specifically for the five dimensional case. In fact $Z(0) = 1 + \frac{\sqrt{c_1}}{4}$.  Generally it is not acceptable to set integration constants to vanish as their values must be settled when matching across a common boundary interface of the interior and exterior spacetime. Note that switching off the higher curvature effects by setting $\alpha = 0$ is not possible.  What is true from (\ref{101}) is that   expansion in powers of $\alpha$ will give the interior Schwarzschild metric of standard Einstein theory as $\alpha \rightarrow 0$ and for arbitrary $c_1$. It shall be seen later that the absence of a singularity in the metric at the centre of an incompressible sphere in EGB theory is only peculiar to the 5-$d$ case and not the 6-$d$ constant density sphere. These results challenge the conclusion that the Schwarzschild metric is universal in EGB theory. Clearly the potential (\ref{101}) generalises the Schwarzschild potential. 

In what follows we shall analyse the impact of imposing conformal flatness on the field equations without any further restrictions. It will be interesting to see whether the interior Schwarzschild metric is preserved  and whether a constant density is an unavoidable consequence of conformal flatness in EGB theory. 

\section{Conformal flatness}\label{sec5}

Conformally flat spacetimes are distinguished by  the vanishing of the Weyl tensor which in turn determines the constraint equation 
\be{r^2(\nu'' + 2\nu'^2 -2\nu' \lambda') - r(\nu' - \lambda') + (1-e^{2\lambda}) = 0}\ee 
for the line element (\ref{5}). This may be converted to the form 
\begin{equation}
4x^2 Z \ddot{y} + 2x^2 \dot{Z}\dot{y} - (\dot{Z}x -Z+1)y = 0, \label{110}
\end{equation}
in our coordinates and must be satisfied for spherically symmetric spacetimes of any dimension.   Equation (\ref{110}) admits the separation of variables
\begin{equation}
y = A\sqrt{x} \cosh \left(\frac{1}{2} \int \frac{dx}{x\sqrt{Z}} + B\right), \label{111}
\end{equation}
where $A$ and $B$ are integration constants. 
Plugging the result (\ref{111}) into the pressure isotropy condition (\ref{8c}) gives the neat factorisation 
\begin{equation}
\left(x \dot{Z}-Z+1\right) \left(-2 \beta  \sqrt{Z} \sinh f(x) +(\beta +3 x) \cosh f(x) -3 \beta  Z \cosh f(x) \right)  =0, \label{112}
\end{equation}
where $f(x) = \left(B+\frac{1}{2} \int \frac{dx}{x \sqrt{Z(x)}} \right)$. The vanishing of the first bracket gives
\begin{equation}
\dot{Z}x -Z + 1 = 0,
\end{equation}
which is solved by $Z = 1 + cx$, $c$ some constant, in general. It can easily be checked that this is the well known Schwarzschild interior solution which yields a constant density or incompressible perfect fluid hypersphere. In standard  Einstein gravity it is known that the Schwarzschild interior metric is conformally flat and conversely demanding that a spacetime be conformally flat yields the Schwarzschild solution. Note that on setting $\beta = 0$ in (\ref{112}) it can easily be checked that Schwarzschild remains the unique solution for 5D Einstein gravity and indeed for any dimension thereafter. However the Schwarzschild solution is not the unique conformally flat solution in the context of Einstein-Gauss-Bonnet gravity.     The second pair of brackets in (\ref{112}) gives the relationship
\begin{equation}
\tanh f(x) = \frac{3x + \beta (1-3Z)}{2\beta \sqrt{Z}} \label{113}
\end{equation}
on rearranging. To solve equation (\ref{113}) for $Z$ in its present form would be difficult so we endeavor to cast it as a differential equation and profit from results on exact solutions of differential equations.
Differentiating (\ref{113}) permits us to recast the equation as
\begin{equation}
\beta  (\beta +3 \beta  Z+3 x) \dot{Z}+\sqrt{Z} \left(-(\beta +3 x)^2-9 \beta ^2 Z^2-6 \beta  \sqrt{Z}+2 \beta  (5 \beta +9 x) Z\right) = 0, \label{114}
\end{equation}
where we used the result $\frac{d(\tanh f)}{dx} = 1 - \tanh ^2 f$. The general solution of equation (\ref{114}) is given by 
\be
Z = \frac{(9x+5\beta)(e^{2x}-k)^2 + 8k\beta e^{2x} \pm 2(e^{2x} +k)\sqrt{\beta} \sqrt{(9x +4\beta)(e^{2x}-k)^2 +4k\beta e^{2x}}}{9\beta (e^{2x}-k)^2}, \label{115}
\ee
where $k$ is a constant of integration.  It is easily checked that the metric potential remains finite at the stellar centre $x = 0$ however this is not the case for the dynamical quantities as shall be seen below. 
In order to establish the function $y$ we use the form (\ref{111}) however, in light of the complexity of the spatial potential $Z$, an explicit form of $y$ free of integrals could not be realised. Nevertheless, the problem is exacerbated by the presence of two square roots in the integrand. It may still be possible to study the physical characteristics of the model generated by the metric with the help of numerical methods.

Now it remains to find the physical dynamics that should be analyzed for their contributions to the 5 dimensional  model. To pursue this we plug (\ref{111}) and (\ref{115}) into the system (\ref{8a})--(\ref{8b}) then find succinct forms for the  energy density and isotropic pressure as 
\begin{eqnarray}
\frac{\rho}{C}&=&\frac{\sqrt{\beta } \left(e^{4 x}-k^2\right) K_8+\left(e^{2 x}-k\right) \left(9 k^2-2 k e^{2 x} (2 \beta +18 x+9)+e^{4 x} (20 \beta +36 x+9)\right)-8 \sqrt{\beta } e^{2 x} J_1}{3 \beta  \left(e^{2 x}-k\right)^3} \nonumber \\
&&+\frac{\sqrt{\beta }\left(e^{2 x}+k\right) K_6 J_2-36 e^{2 x}\left(e^{2 x}-k\right)^2 \left(\left(k^2+e^{4 x}\right) (5 \beta +9 x)+2 k e^{2 x} (\beta +9 x)\right)J_1}{27 \sqrt{\beta } x \left(e^{2 x}-k\right)^5 J_1}, \label{116}  \\ 
\frac{p}{C}&=& \frac{K_4 -6\beta\left(e^{2 x}-k\right)  \left(K_1-2 \sqrt{\beta } \left(k^2-4 k e^{2 x}+e^{4 x}\right)\right) \tanh \left(B+\frac{3}{2} K_3\right) \sqrt{K_2+2\sqrt{\beta}K_1}}{27 \beta  x^2 \left(e^{2 x}-k\right)^4},\label{117}
\end{eqnarray}
where for ease of reference we put \\ 
$
J_1=\sqrt{\left(k+e^{2 x}\right)^2 \left(k^2 (4 \beta +9 x)-2 k e^{2 x} (2 \beta +9 x)+e^{4 x} (4 \beta +9 x)\right)}  
$,\\
$
J_2=9 k^4-16 k^2 e^{4 x} (\beta +9 x)+2 k e^{2 x} \left(k^2 (20 \beta +36 x-9)+8 \sqrt{\beta } J_1\right)+2 k e^{6 x} (20 \beta +36 x+9)-9 e^{8 x}$,\\
$
K_1=\sqrt{k^4 (4 \beta +9 x)+4 \beta  k^3 e^{2 x}-18 k^2 e^{4 x} x+4 \beta  k e^{6 x}+e^{8 x} (4 \beta +9 x)}$,\\
$K_2=e^{4 x} (5 \beta +9 x)-2 k e^{2 x} (\beta +9 x)+k^2 (9x+5\beta)$, \\ 
$K4=k^4 (9 x-2 \beta )^2-2 \beta ^{3/2} k^2 K_1+2 e^{4 x} \left(9 k^2 \left(4 \beta ^2+27 x^2+4 \beta  x\right)-\beta ^{3/2} K_1\right)-4 k e^{2 x} \left(k^2 \left(26 \beta ^2+81 x^2\right)+7 \beta ^{3/2} K_1\right)-4 k e^{6 x} \left(26 \beta ^2+81 x^2\right)+e^{8 x} (9 x-2 \beta )^2$,\\
$K_5K_1=9 k^4+8 \beta  k^3 e^{2 x}-18 k^2 e^{4 x}-72 k^2 e^{4 x} x+24 \beta  k e^{6 x}+8 e^{8 x} (4 \beta +9 x)+9 e^{8 x}$,\\
$K_6=2 \sqrt{\beta } J_1-4 \beta  k^2+9 k^2 x-2 k e^{2 x} (9 x-8 \beta )+e^{4 x} (9 x-4 \beta )$,\\
$K_8J_1=9 k^3+(8 \beta -9) k^2 e^{2 x}-k e^{4 x} (8 \beta +72 x+9)+e^{6 x} (32 \beta +72 x+9)$,\\
$ K_3 =\int \left(x \sqrt{\frac{K_2 +2\sqrt{\beta}J_1}{\beta  \left(e^{2 x}-k\right)^2}} \, \right)^{-1} \, dx$. \\
Invoking (\ref{116}) and (\ref{117}) renders the ratio
\beq
S=\frac{dp}{d\rho}&=&\left[\frac{8 e^{4 x} \left(9 k^2 \left(27 x^2+4 \beta  x+4 \beta ^2\right)-{K_1} \beta ^{3/2}\right)-24 e^{6 x} k \left(81 x^2+26 \beta ^2\right)-8 e^{2 x} k \left(7 K_1 \beta ^{3/2}+k^2 \left(81 x^2+26 \beta ^2\right)\right)}{27 \left(e^{2 x}-k\right)^4 x^2 \beta } \right. \n \\
&&\left.+\frac{2 \left(9 k^2 (54 x+4 \beta )-k^2 K_5 \beta ^{3/2}-\frac{1}{2} \beta ^{3/2} K_5\right) e^{4 x}+8 e^{8 x} (9 x-2 \beta )^2-4 e^{2 x} k \left(\frac{7}{2} K_5 \beta ^{3/2}+162 k^2 x\right)-648 e^{6 x} k x}{27 \left(e^{2 x}-k\right)^4 x^2 \beta } \right. \n \\
&&\left.+\frac{8 e^{2 x} K_9 \tanh \left(B+\frac{3 }{2}K_3\right)- \sqrt{2 \sqrt{\beta } K_1+K_2} \left(K_5-4 \left(4 e^{4 x}-8 e^{2 x} k\right) \sqrt{\beta }\right) \tanh \left(B+\frac{3 }{2}K_3\right)}{9 \left(e^{2 x}-k\right)^3 x^2}-\frac{8 e^{2 x} K_4}{27 \left(e^{2 x}-k\right)^5 x^2 \beta } \right.\n \\
&&\left.-\frac{K_9 \beta^{3/2}  \text{sech}^2\left(B+\frac{3}{2}K_3\right)}{3 \left(e^{2 x}-k\right)x^3 \sqrt{2 \sqrt{\beta } J_1+K_2}}-\frac{2 K_4-18 e^{8 x} (9 x-2 \beta )-18 k^4 (9 x-2 \beta )}{27 \left(e^{2 x}-k\right)^4 x^3 \beta } \right. \n \\
&&\left.+\frac{4 K_9 \tanh \left(B+\frac{3 }{2}K_3\right)}{9 \left(e^{2 x}-k\right)^2 x^3}-\frac{K_9 \beta ^{3/2}K_{10}  \tanh \left(B+\frac{3 }{2}K_3\right)}{9 x^2 \left(2 \sqrt{\beta } K_1+K_2 \right)} \right] \div \n \\
&&\left[\frac{2J_2^3 \left(k+e^{2 x}\right) \left(K_{11}+K_7 \sqrt{\beta }\right)-J_2 \left(k+e^{2 x}\right)^2 4 e^{2 x}K_6 \left((9 x+4 \beta ) k^2-2 e^{2 x} (9 x+2 \beta ) k+e^{4 x} (9 x+4 \beta )\right) }{54 \left(e^{2 x}-k\right)^5 x \sqrt{\beta } J_1^{3}} \right. \n \\
&&\left.+\frac{2 e^{2 x} \left(9 k^2-2 e^{2 x} (18 x+2 \beta +9) k+e^{4 x} (36 x+20 \beta +9)\right)-8 e^{2 x} \left(\left(k^2+e^{4 x}\right) (9 x+5 \beta )-2 e^{2 x} k (9 x+\beta )\right)}{3 \left(e^{2 x}-k\right)^3 \beta } \right. \n \\
&&\left.\frac{+2 e^{2 x} \left(9 k^2-2 e^{2 x} (18 x+2 \beta +9) k+e^{4 x} (36 x+20 \beta +9)\right)+2 e^{2 x} \left(e^{2 x}-k\right) K_8 \sqrt{\beta }+2 e^{2 x} \left(k+e^{2 x}\right) K_8 \sqrt{\beta }}{3 \left(e^{2 x}-k\right)^3 \beta } \right.\n \\
&&\left.\frac{+\left(e^{2 x}-k\right) \left(36 e^{4 x}-36 e^{2 x} k-4 e^{2 x} (18 x+2 \beta +9) k+4 e^{4 x} (36 x+20 \beta +9)\right)}{3 \left(e^{2 x}-k\right)^3 \beta }-\frac{J_2 \left(k+e^{2 x}\right)^3 K_6 K_{11}}{54 \left(e^{2 x}-k\right)^5 x \sqrt{\beta } J_1^{3}} \right. \n \\
&&\left.+\frac{\left(k+e^{2 x}\right) K_6 \left(72 e^{6 x} k-144 e^{4 x} k^2-64 e^{4 x} (9 x+\beta ) k^2+2 e^{2 x} \left(36 k^2+4 K_7 \sqrt{\beta }\right) k \right)-J_2 \left(k+e^{2 x}\right) K_6}{27 J_1 \left(e^{2 x}-k\right)^5 x \sqrt{\beta }} \right.\n \\
&&\left.+\frac{\left(k+e^{2 x}\right) K_6\left(12 e^{6 x} (36 x+20 \beta +9) k+4 e^{2 x} \left((36 x+20 \beta -9) k^2+8 J_1 \sqrt{\beta }\right) k-72 e^{8 x}\right)+2 e^{2 x} J_2 K_6}{27 J_1 \left(e^{2 x}-k\right)^5 x \sqrt{\beta }} \right.\n \\
&&\left.+\frac{\left(e^{4 x}-k^2\right) \left(2 e^{2 x} (8 \beta -9) k^2-72 e^{4 x} k-4 e^{4 x} (72 x+8 \beta +9) k+72 e^{6 x}+6 e^{6 x} (72 x+32 \beta +9)\right)}{3 \left(e^{2 x}-k\right)^3 \sqrt{\beta }  J_1} \right.\n \\
&&\left.+\frac{2 e^{2 x} \left(8 e^{2 x} \sqrt{\beta } J_1+4 e^{2 x} \left(\left(k^2+e^{4 x}\right) (9 x+5 \beta )+2 e^{2 x} k (9 x+\beta )\right)\right)}{\left(e^{2 x}-k\right)^4 \beta }-\frac{10 e^{2 x} J_2\left(k+e^{2 x}\right) K_6}{27 J_1 \left(e^{2 x}-k\right)^6 x \sqrt{\beta }}\right.  \n \\
&&\left.+\frac{4 e^{4 x} (9 x+5 \beta )-16 e^{2 x} \sqrt{\beta } J_1-4 e^{2 x} \sqrt{\beta } K_7-4 e^{2 x} \left(9 \left(k^2+e^{4 x}\right)-18 e^{2 x} k-4 e^{2 x} (9 x+\beta ) k \right)}{3 \left(e^{2 x}-k\right)^3 \beta } \right. \n \\
&&\left.-\frac{2 e^{2 x}\left(e^{2 x}-k\right) \left(9 k^2-2 e^{2 x} (18 x+2 \beta +9) k+e^{4 x} (36 x+20 \beta +9)\right)+\left(e^{4 x}-k^2\right)K_8 \sqrt{\beta }}{\left(e^{2 x}-k\right)^4 \beta } \right. \n \\
&&\left.-\frac{\left(e^{4 x}-k^2\right)  \left(9 k^3+e^{2 x} (8 \beta -9) k^2-e^{4 x} (72 x+8 \beta +9) k+e^{6 x} (72 x+32 \beta +9)\right) K_7}{6 \left(e^{2 x}-k\right)^3 \sqrt{\beta } J_1^2}\right],
\eeq
which represents the square of the speed of sound and where we  make the additional  substitutions  
\beq
K_7J_1&=&\sqrt{\beta } \left(\left(k+e^{2 x}\right)^2 \left(9 k^2-4 k e^{2 x} (2 \beta +9 x)-18 k e^{2 x}+4 e^{4 x} (4 \beta +9 x)+9 e^{4 x}\right) \right. \n \\
&&\left.+4 e^{2 x} \left(k+e^{2 x}\right) \left(k^2 (4 \beta +9 x)-2 k e^{2 x} (2 \beta +9 x)+e^{4 x} (4 \beta +9 x)\right)\right), \n \\
K_8J_1&=&9 k^3+(8 \beta -9) k^2 e^{2 x}-k e^{4 x} (8 \beta +72 x+9)+e^{6 x} (32 \beta +72 x+9), \n \\
K_9&=&\left(K_1-2 \left(k^2-4 e^{2 x} k+e^{4 x}\right) \sqrt{\beta }\right) \sqrt{\frac{2 \sqrt{\beta } K_1+K_2}{\left(e^{2 x}-k\right)^2 \beta }}, \n \\
K_{10}&=&\frac{9 k^2-18 e^{2 x} k-4 e^{2 x} (9 x+\beta ) k+9 e^{4 x}+4 e^{4 x} (9 x+5 \beta )+K_5 \sqrt{\beta }}{\left(e^{2 x}-k\right)^2 \beta }-\frac{4 e^{2 x} \left(2 \sqrt{\beta } K_1+K_2\right)}{\left(e^{2 x}-k\right)^3 \beta }, \n \\
K_{11}&=& 9 k^2-18 e^{2 x} k-4 e^{2 x} (9 x+2 \beta ) k+9 e^{4 x}+4 e^{4 x} (9 x+4 \beta ), \n
\eeq
to alleviate the complexity of the expression.
Using our expressions (\ref{116}) and (\ref{117}) we are now able to find expressions for the  weak, strong and dominant energy conditions in the forms 
\beq
\frac{\rho -p}{C}&=&\frac{2 \sqrt{\beta } K_{13}\left(K_1-2 \left(k^2-4 e^{2 x} k+e^{4 x}\right) \sqrt{\beta }\right) \tanh \left(B+\frac{3}{2}K_3\right)}{9 \left(e^{2 x}-k\right)^2 x^2}-\frac{K_{12}}{27 \left(e^{2 x}-k\right)^4 x^2 \beta }+\frac{ J_2K_6\left(e^{2 x}+k\right)}{27 \left(e^{2 x}-k\right)^5 x J_1\sqrt{\beta } } \n \\
&&+\frac{\left(e^{4 x}-k^2\right) \sqrt{\beta } K_8+\left(e^{2 x}-k\right) \left(9 k^2-2 e^{2 x} (18 x+2 \beta +9) k+e^{4 x} (36 x+20 \beta +9)\right)-8 e^{2 x} \sqrt{\beta }J_1-4K_2e^{2 x}}{3 \left(e^{2 x}-k\right)^3 \beta }, \n \\\\ \label{118}
\frac{\rho +p}{C}&=&\frac{2 \sqrt{\beta }K_{13} \left(2 \left(k^2-4 e^{2 x} k+e^{4 x}\right) \sqrt{\beta }-K_1\right) \tanh \left(B+\frac{3}{2}K_3 \right)}{9 \left(e^{2 x}-k\right)^2 x^2}-\frac{4 e^{2 x} \left(\left(k^2+e^{4 x}\right) (9 x+5 \beta )+2 e^{2 x} k (9 x+\beta )\right)}{3 \left(e^{2 x}-k\right)^3 \beta } \n \\
&&+\frac{\left(e^{4 x}-k^2\right) \sqrt{\beta } K_8+\left(e^{2 x}-k\right) \left(9 k^2-2 e^{2 x} (18 x+2 \beta +9) k+e^{4 x} (36 x+20 \beta +9)\right)-8 e^{2 x} \sqrt{\beta }J_1}{3 \left(e^{2 x}-k\right)^3 \beta } \n \\
&&+\frac{K_{12}}{27 \beta \left(e^{2 x}-k\right)^4 x^2 }+\frac{ J_2K_6\left(e^{2 x}+k\right)}{27 \left(e^{2 x}-k\right)^5 x J_1\sqrt{\beta } },
 \label{119} \\
\frac{\rho +4p}{C}&=&\frac{K_{13} \left(2 \left(k^2-4 e^{2 x} k+e^{4 x}\right) \sqrt{\beta }-K_1\right) \tanh \left(B+\frac{3}{2} K_3\right)}{9 \left(e^{2 x}-k\right)^2 x^2}-\frac{4 e^{2 x} \left(\left(k^2+e^{4 x}\right) (9 x+5 \beta )+2 e^{2 x} k (9 x+\beta )\right)}{3 \left(e^{2 x}-k\right)^3 \beta }\n \\
&&+\frac{\left(e^{4 x}-k^2\right) \sqrt{\beta } K_8+\left(e^{2 x}-k\right) \left(9 k^2-2 e^{2 x} (18 x+2 \beta +9) k+e^{4 x} (36 x+20 \beta +9)\right)-8 e^{2 x} \sqrt{\beta }J_1}{3 \left(e^{2 x}-k\right)^3 \beta } \n \\
&& +\frac{4K_{12}}{27 \beta \left(e^{2 x}-k\right)^4 x^2  }+\frac{ J_2K_6\left(e^{2 x}+k\right)}{27 \left(e^{2 x}-k\right)^5 x J_1\sqrt{\beta } },   \label{120} 
\eeq
respectively. We have additionally set\\
$K_{12}=2 e^{4 x} \left(9 k^2 \left(27 x^2+4 \beta  x+4 \beta ^2\right)-\beta ^{3/2} J_1\right)-4 e^{2 x} k \left(7 J_1 \beta ^{3/2}+k^2 \left(81 x^2+26 \beta ^2\right)\right)+e^{8 x} (9 x-2 \beta )^2-2 k^2 J_1 \beta ^{3/2}+k^4 (9 x-2 \beta )^2-4 e^{6 x} k \left(81 x^2+26 \beta ^2\right)$,
 $K_{13}= \sqrt{\frac{K_2+2 \sqrt{\beta }J_1}{\left(e^{2 x}-k\right)^2 \beta }}$ and
$K_{14}=\frac{K_4}{27 \left(e^{2 x}-k\right)^4 x^2 \beta }$.\\
The Chandrasekhar stability index is given by
\begin{eqnarray}
\Gamma &=& \left(e^{2 x}-k\right) \left(\left(e^{2 x}+k\right)K_{15}\right)^{3/2} \left(\frac{\beta  K_{16}}{(9 x+4 \beta ) k^3-9 e^{2 x} x k^2-9 e^{4 x} x k+e^{6 x} (9 x+4 \beta )}\right. \nonumber\\
&& \left. +\frac{3 \left(e^{2 x}-k\right)^2 \beta^{3/2} \sech^2 \left(B+\frac{3 }{2}K_3\right) K_{17}}{2 K_{15}\left(2 \sqrt{\beta} K_1+K_2\right)}\right) \times \left(\frac{K_{12}}{27 \left(e^{2 x}-k\right)^4 x^2 \beta}+\frac{\left(k+e^{2 x}\right) K_6 K_{18}}{27 \left(e^{2 x}-k\right)^5 x \sqrt{\beta} J_1} \right. \nonumber\\
&& \left. +\frac{2 \left(2 \left(k^2-4 e^{2 x} k+e^{4 x}\right) \sqrt{\beta}-K_1\right) \sqrt{2 \sqrt{\beta} J_1+K_2 } \tanh \left(B+\frac{3 }{2}K_3\right)}{9 \left(e^{2 x}-k\right)^3 x^2} \right. \nonumber\\
&& \left. +\frac{\left(e^{2 x}-k\right) K_{19}-8 e^{2 x} \sqrt{\beta } J_1-4 e^{2 x} K_2+\left(e^{4 x}-k^2\right)  K_8 \sqrt{\beta }}{3 \left(e^{2 x}-k\right)^3 \beta }\right) \div \nonumber\\
&&\left[\frac{1}{27 \left(e^{2 x}-k\right)^4 x^2 \beta  }\left(2 \left(k+e^{2 x}\right)^2 x \beta  \left(9 \left((9 x+4 \beta ) J_1-k^2 \sqrt{\beta } (27 x+8 \beta )\right) k^8-36\beta k^7 e^{2 x}\left(20 x^2+6 x+3\right)J_1 \right.\right.\right. \nonumber \\
&&\left.\left.\left.-4 e^{2 x} \left(\left(8\beta ^2 (2 x-1) +81 x \left(4 x^2+2 x+1\right)\right) J_1-2 k^2 \sqrt{\beta } \left(8 (2 x-1) \beta ^2+9 (4 (x-1) x+7) \beta +243 x\right)\right) k^7 \right.\right.\right.\n \\
&&\left.\left.\left.-e^{4 x} \left(\sqrt{\beta } \left(64 (95 x-27) \beta ^2+72 (64 x (6 x-1)+15) \beta +243 x \left(128 x^2+21\right)\right) k^2+4 J_1 \left(648 (2 \beta -1) x^2\right)\right) k^6 \right.\right. \right.\nonumber \\
&&\left.\left. \left.+4 e^{6 x} \left(4 \sqrt{\beta } \left(-16 (19 x+11) \beta ^2+36 (x (28 x-17)+1) \beta +243 \left(16 x^3+x\right)\right) k^2+36 \beta \left(20 x^2-78 x+3\right)J_1  \right.\right.\right.\right.\nonumber \\
&&\left.\left.\left.\left.+J_1 \left(-8 (114 x+13) \beta ^2 +81 x (2 x (18 x-5)+1)\right)\right) k^5 +2\sqrt{\beta } e^{8 x} \left(32 (43-65 x) \beta ^2+576 (6 x (4 x+1)+1) \beta\right)k^2\right.\right. \right.\nonumber \\
&&\left.\left. \left.+2 e^{8 x} \left(\sqrt{\beta } \left( 243 x \left(64 x^2+11\right)\right) k^2+J_1 (x (-64 (9 x+\beta ) (18 x-11 \beta )-405)-108 \beta )\right) k^4 +32\beta ^2k^3 e^{10 x}(13-114 x)  \right.\right.\right. \nonumber \\
&&\left.\left.\left.+4 e^{10 x} \left(\left(9 \left(20 x^2+78 x+3\right) \beta +81 x (2 x (18 x+5)+1)\right) J_1-4 k^2 \sqrt{\beta } \left(27 \left(76 x^2+5\right) \beta +243 x \left(32 x^2+3\right)\right)\right) k^3 \right.\right. \right.\nonumber \\
&&\left.\left.\left.-2 e^{12 x} \left(\sqrt{\beta } \left(32 (65 x+43) \beta ^2-576 (6 x (4 x-1)+1) \beta -243 x \left(64 x^2+11\right)\right) k^2+2k^2(48 \beta  (11 \beta +9)-81) x J_1 \right.\right. \right. \right.\nonumber \\
&&\left.\left.\left.\left.+2 J_1 \left(648 (2 \beta +1) x^2+2 \beta  (52 \beta -9)\right)\right) k^2+16\sqrt{\beta }k^2 e^{14 x}\left(16 (11-19 x) \beta ^2+36 (x (28 x+17)+1) \beta\right) \right. \right.\right.\nonumber \\
&&\left.\left.\left.+4 e^{14 x} \left(4 k^2 \sqrt{\beta } \left(243 \left(16 x^3+x\right)\right)-\left(8 (2 x+1) \beta ^2+9 \left(20 x^2-6 x+3\right) \beta +81 x \left(4 x^2-2 x+1\right)\right) J_1\right) k\right.\right. \right.\nonumber \\
&&\left.\left.\left.+8 e^{18 x} \sqrt{\beta } \left(8 (2 x+1) \beta ^2+9 (4 x (x+1)+7) \beta +243 x\right) k-e^{4x}\left((48 \beta  (11 \beta -9)-81) x-2 \beta  (52 \beta +9)\right)J_1 \right. \right.\right. \nonumber \\
&&\left.\left.\left.+e^{16 x} \left(9 (9 x+4 \beta ) J_1-k^2 \sqrt{\beta } \left(64 (95 x+27) \beta ^2+72 (64 x (6 x+1)+15) \beta +243 x \left(128 x^2+21\right)\right)\right)\right.\right. \right.\nonumber \\
&&\left.\left. \left. -11008k^5 \beta ^{5/2} x e^{10x} -9 e^{20 x} \sqrt{\beta } (27 x+8 \beta )\right) K_4 \right)-\frac{2 \sqrt{\beta } K_9 \tanh \left(B+\frac{3 }{2}K_3\right)}{9 \left(e^{2 x}-k\right)^2 x^2}\right],
\end{eqnarray}
where
\begin{eqnarray}
K_{15}&=&\left(k+e^{2 x}\right) \left((9 x+4 \beta ) k^2-2 e^{2 x} (9 x+2 \beta ) k+e^{4 x} (9 x+4 \beta )\right), \nonumber \\
K_{16}&=& 8k^7e^{2x}\left(81 (4 x+1) x^2+18 (19 x-13) \beta  x+4 (22 x-27) \beta ^2\right)-\left(4 (9 x+4 \beta ) (9 x-2 \beta ) k^2+\sqrt{\beta } (27 x+16 \beta ) K_1\right) k^6 \nonumber \\
&&+4 e^{2 x} \left(2 \left(81 (4 x+1) x^2+18 (19 x-13) \beta  x+4 (22 x-27) \beta ^2\right) k^2+K_1 \sqrt{\beta } (x (162 x+74 \beta -81)-52 \beta )\right) k^5 \n \\
&&-e^{4 x} \left(8 k^2 \left(81 (8 x-1) x^2-18 (2 x+37) \beta  x-8 (21 x+22) \beta ^2\right)-\sqrt{\beta } K_1 (208 \beta +x (864 x+800 \beta +729))\right) k^4 \nonumber \\
&&-8 e^{6 x} \left(\left(81 (4 x+3) x^2+18 (19 x+33) \beta  x+4 (43-42 x) \beta ^2\right) k^2+18 K_1 x \sqrt{\beta } (21 x+\beta )\right) k^3 \n \\
&&+e^{8 x} \left(64 x \left(162 x^2-9 \beta  x+22 \beta ^2\right) k^2+K_1 \sqrt{\beta } (x (864 x+800 \beta -729)-208 \beta )\right) k^2+4 e^{16 x} (9 x+4 \beta ) (9 x-2 \beta ) \nonumber \\
&&+8 e^{14 x} \left(81 (4 x-1) x^2+18 (19 x+13) \beta  x+4 (22 x+27) \beta ^2\right) k+4ke^{10 x}\sqrt{\beta } K_1 (52 \beta +x (162 x+74 \beta +81)) \nonumber \\
&&-4 e^{10 x} \left(2 k^2 \left(81 (4 x-3) x^2+18 (19 x-33) \beta  x-4 (42 x+43) \beta ^2\right)\right) k+\sqrt{\beta }e^{12 x} (27 x+16 \beta ) K_1 \nonumber \\
&&+e^{12 x} \left(-8 k^2 \left(81 (8 x+1) x^2+18 (37-2 x) \beta  x+8 (22-21 x) \beta ^2\right)\right), \nonumber \\
K_{17}&=&6 e^{4 x} \left(e^{-2 x} k^2-e^{2 x}\right) \left(-2 e^{8 x} \beta ^{3/2}+4 e^{6 x} k (27 x+13 \beta ) \sqrt{\beta }+k^2 \left((9 x+\beta ) K_1-2 k^2 \beta ^{3/2}\right)\right)-36 k^2 \sqrt{\beta }e^{4 x} (6 x+\beta ) \nonumber \\
&&+e^{4 x} \left((9 x+\beta ) K_1+2 e^{2 x} k \left(2 \sqrt{\beta } (27 x+13 \beta ) k^2+K_1 (7 \beta -9 x)\right)\right) \left(e^{-2 x} \left(k^2+e^{4 x}\right) (9 x+4 \beta )-2 k (9 x+2 \beta )\right) \n \\
&&+\left(\left(k^2 \sqrt{\beta } (9 x+4 \beta ) (9 x+8 \beta )-\left(162 x^2+117 \beta  x+16 \beta ^2\right) K_1\right) k^6+4 K_1 e^{2 x}\left(9 (28 x-3) \beta  x+26 (3 x-2) \beta ^2\right)k^5 \right.\nonumber \\
&&\left.+2 e^{2 x} \left(3 \sqrt{\beta } \left(27 (20 x-19) x^2+6 (82 x-93) \beta  x+16 (7 x-9) \beta ^2\right) k^2+2 K_1 \left(162 (x+1) x^2\right)\right) k^5 \right. \nonumber \\
&&\left.-e^{4 x} \left(\left(162 (16 x+5) x^2+27 (32 x-21) \beta  x-16 (24 x+13) \beta ^2\right) K_1-2 k^2 \sqrt{\beta } \left(81 (71-40 x) x^2\right)\right) k^4 \right. \nonumber \\
&&\left.+2 e^{6 x} \left(\sqrt{\beta } \left(-81 (20 x+87) x^2-18 (82 x+225) \beta  x+16 (27 x-43) \beta ^2\right) k^2+72 K_1 x \left(27 x^2-2 \beta  x+\beta ^2\right)\right) k^3 \right. \nonumber \\
&&\left.+e^{8 x} \left(96 k^2 x \sqrt{\beta } \left(135 x^2+21 \beta  x+14 \beta ^2\right)-\left(162 (16 x-5) x^2+27 (32 x+21) \beta  x+16 (13-24 x) \beta ^2\right) K_1\right) k^2 \right. \nonumber \\
&&\left.+2 e^{10 x} \left(\sqrt{\beta } \left(81 (87-20 x) x^2+18 (225-82 x) \beta  x+16 (27 x+43) \beta ^2\right) k^2+2 K_1 \left(162 (x-1) x^2+9 (28 x+3) \beta  x\right)\right) k \right. \nonumber \\
&&\left.+6 e^{14 x} \sqrt{\beta } \left(27 (20 x+19) x^2+6 (82 x+93) \beta  x+16 (7 x+9) \beta ^2\right) k +72k^6\beta^{3/2}e^{4x} (123-14 x)  x\right. \nonumber \\
&&\left.+e^{12 x} \left(\left(162 x^2+117 \beta  x+16 \beta ^2\right) K_1-2 k^2 \sqrt{\beta } \left(81 (40 x+71) x^2+36 (14 x+123) \beta  x+16 (44-27 x) \beta ^2\right)\right) \right. \nonumber \\
&&\left.+104k\beta^2 (3 x+2)K_1-32 k^6 \beta ^{5/2}e^{4x} (27 x+44) -e^{16 x} \sqrt{\beta } (9 x+4 \beta ) (9 x+8 \beta )\right) \sinh (2 B+3 K_3) \sqrt{\frac{K_2+2 K_1 \sqrt{\beta }}{\left(e^{2 x}-k\right)^2 \beta }}, \nonumber \\
K_{18}&=&9 k^4-16 e^{4 x} (9 x+\beta ) k^2+2 e^{2 x} \left((36 x+20 \beta -9) k^2+8 J_1 \sqrt{\beta }\right) k+2 e^{6 x} (36 x+20 \beta +9) k-9 e^{8 x}, \nonumber \\
K_{19}&=&9 k^2-2 e^{2 x} (18 x+2 \beta +9) k+e^{4 x} (36 x+20 \beta +9). \nonumber 
\end{eqnarray}  
Inserting the  energy density profile into the  gravitational mass formula gives 
\beq
M&=&\frac{1}{3\sqrt{\beta }C^{3/2}}\int x^{3/2} \left(\frac{e^{4 x} (20 \beta +36 x+9)-8 \sqrt{\beta } J_1 e^{2 x}+\left(e^{2 x}-k\right) \left(9 k^2-2 k e^{2 x} (2 \beta +18 x+9)\right)}{\sqrt{\beta } \left(e^{2 x}-k\right)^3} \right. \n \\
&&\left.+\frac{ K_8 \sqrt{\beta }\left(e^{2 x}+k\right)-4 K_2 e^{2 x}}{\sqrt{\beta } \left(e^{2 x}-k\right)^2}+\frac{\left(e^{2 x}+k\right) J_2 K_6 }{9 \left(e^{2 x}-k\right)^5xJ_1}\right)  \, dx ,
\eeq
for a $5-d$ hypersphere of perfect fluid. 

\begin{figure}[h]
\centering
\includegraphics[width=0.43\linewidth]{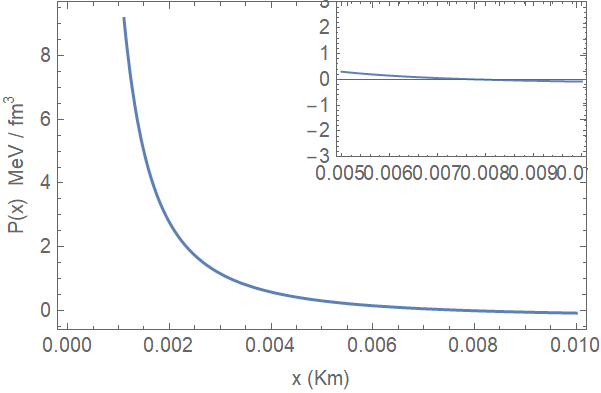}
\includegraphics[width=0.43\linewidth]{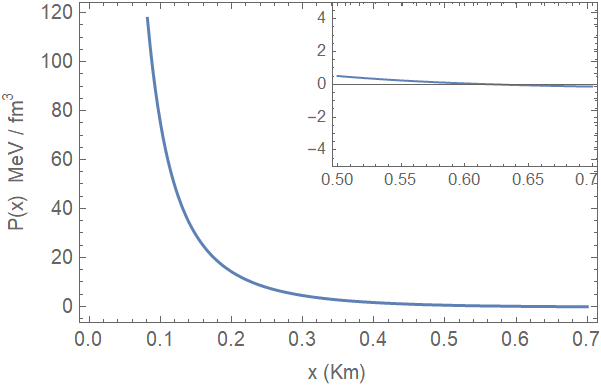}
\caption{The pressure profile is plotted as a function of radius for $5D$ and $6D$ EGB gravity, respectively. 
 } \label{f1}
\end{figure}

\begin{figure}[h]
\centering
\includegraphics[width=0.43\linewidth]{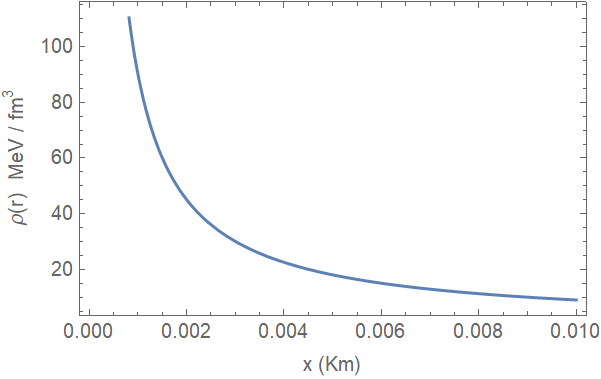}
\includegraphics[width=0.43\linewidth]{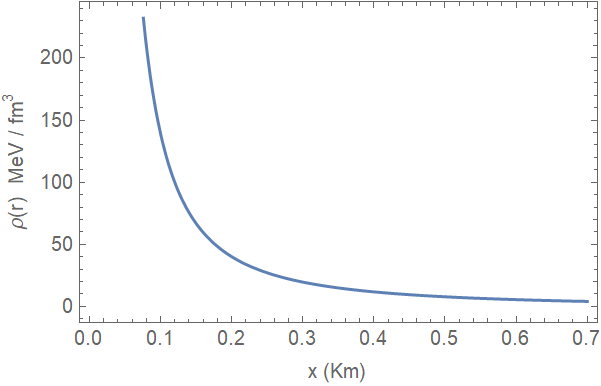}
\caption{Variation of density as a function of the radial
coordinate for compact star. The labels of the curves are the same as given in Fig. \ref{f1}.} \label{f2}
\end{figure}

\begin{figure}[h]
\centering
\includegraphics[width=0.43\linewidth]{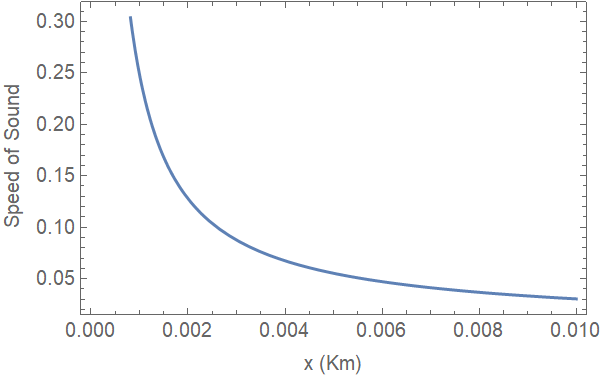}
\includegraphics[width=0.43\linewidth]{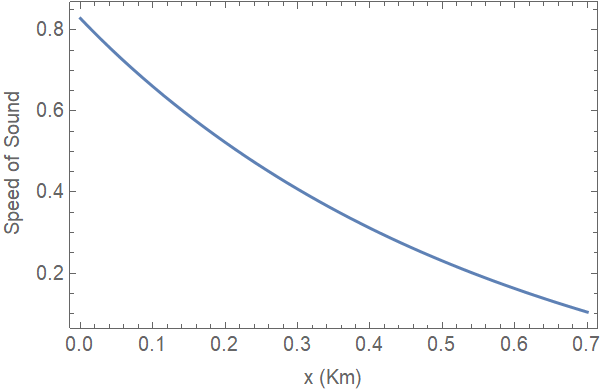}
\caption{Variation of speed of sound with radius. } \label{f3}
\end{figure}
\begin{figure}[h]
\centering
\includegraphics[width=0.43\linewidth]{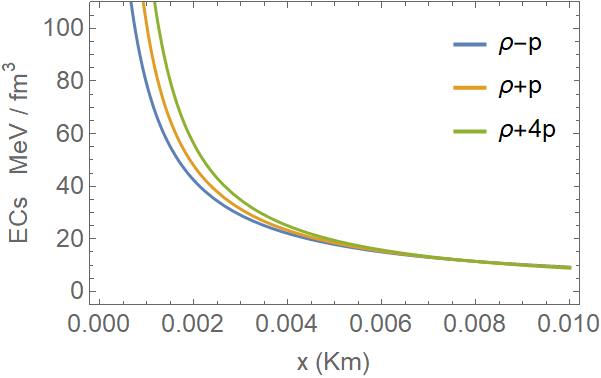}
\includegraphics[width=0.43\linewidth]{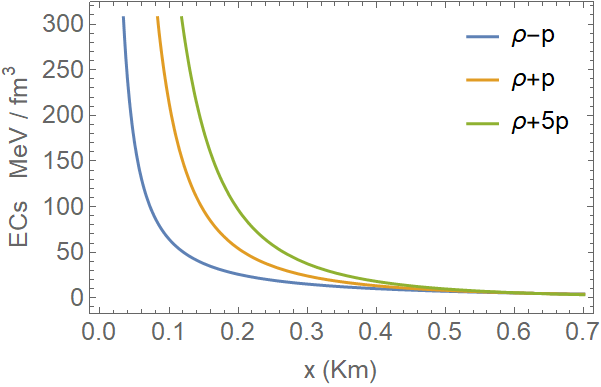}
\caption{Variation of energy conditions with radius. } \label{f4}
\end{figure}

\begin{figure}[h]
\centering
\includegraphics[width=0.43\linewidth]{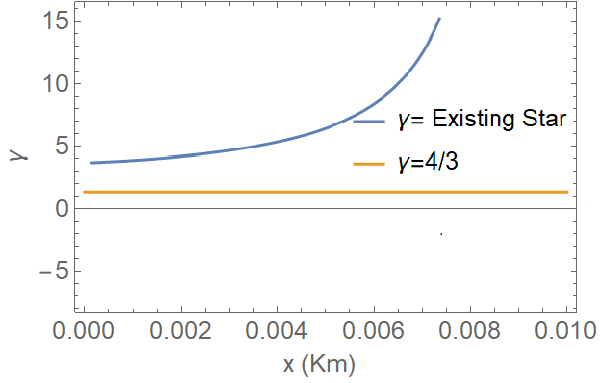}
\includegraphics[width=0.45\linewidth]{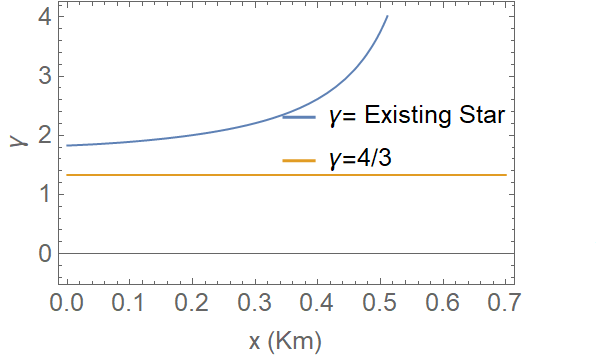}
\caption{Variation of adiabatic index with radius. } \label{f5}
\end{figure}

\begin{figure}[h]
\centering
\includegraphics[width=0.43\linewidth]{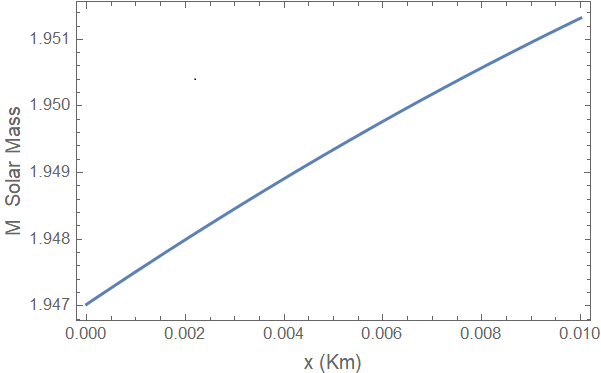}
\includegraphics[width=0.43\linewidth]{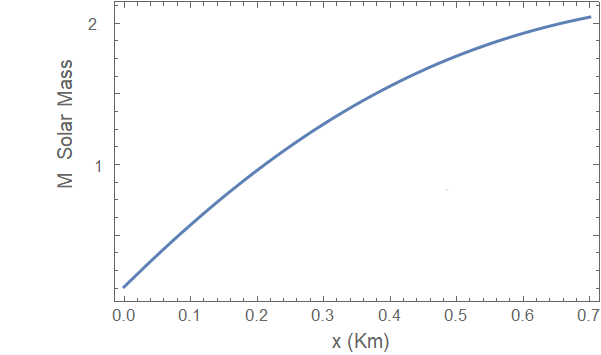}
\caption{Mass as a function of the radial
coordinate for compact star.  In both
cases the parameters are given in Fig. \ref{f1}.  } \label{f6}
\end{figure}

\begin{figure}[h]
\centering
\includegraphics[width=0.49\linewidth]{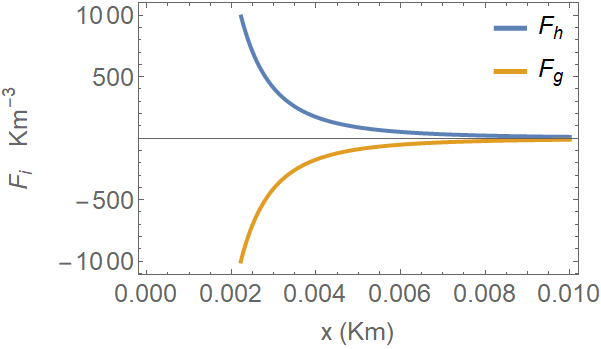}
\includegraphics[width=0.45\linewidth]{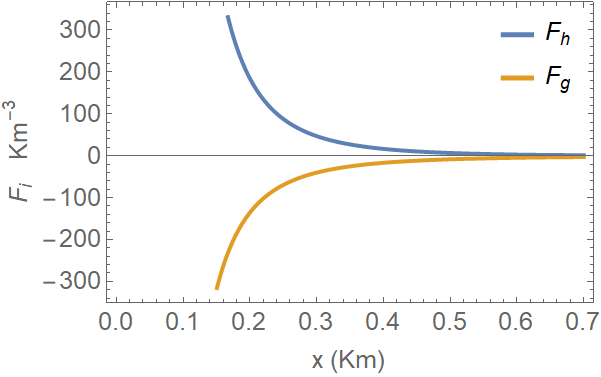}
\caption{Variation of hydrodynamic ($F_h$) and gravitational ($F_g$) forces versus the radial coordinate $x$.   } \label{f9}
\end{figure}

\subsection{Physical admissibility of the five dimensional solution}

We now proceed to analyse in a qualitative sense the physical properties of our new conformally flat spacetime in EGB. The very complicated and lengthy expressions that emanate prohibit an analytical treatment of the physics therefore we have resorted to a graphical analysis using suitably selected parameter spaces.  
The physical characteristics of the 5-$d$ hypersphere are displayed in the left panels of each of the figures while the 6-d plots appear on the right. In the case of the 5-d plots we have selected the parameters as follows after extensive empirical fine-tuning:   $A=4, ~B=1,~C=0.06, k=0.6,~ \beta =0.4$ and $\alpha =0.01$. 

It is worth noting at this point that  the specific models we are presenting are only valid for the choices of parameter space in both 5 and 6 dimensions. There are a number of parameters that are moving parts in the system so it is not possible to say that the solution is generic. We  have exhibited the existence of at least one physically well behaved model for the general EGB spacetime we have found in 5 and 6D. Researchers working in simulations may utilise our general solution and incorporate data pertaining to known structures to make an identification with observed phenomena such as neutron stars, white dwarfs, cold planets and so on.

\begin{enumerate}
    \item [(i)] From Fig. \ref{f1}  it may be  observed that the pressure decreases  smoothly towards the surface layer of the star. The pressure vanishes for some finite radius corresponding approximately to $x = 0.01$ which defines the boundary of the star. This immediately suggests that the model may represent a compact star.  \\
    \item[(ii)] The density profile  plotted as a function of the radial coordinate $x$ in Fig. \ref{f2}  reveals that  the density is a monotonically decreasing function as one moves from the stellar center towards the stellar boundary. This is considered to be physically reasonable.   \\
    \item[(iii)] In order to prevent superluminal sound  speeds within the stellar fluid it is required that the condition $0 \leq \frac{dp}{d\rho} \leq 1$ be satisfied throughout the stellar interior.   From Fig. \ref{f3} we observe that this criterion is met everywhere except at the centre of the star. In order for our model to obey causality requirements we must make an incision from $r = 0$ to some finite radius $r = r_0$, thus removing this portion of the solution. We can then replace this portion with a well--behaved solution from the center ($r = 0$) to $r = r_0$ and match it to our present solution from $r = r_0$ to $r = R$ where $R$ is the boundary of the star. In this way we can effectively create core-envelope models of compact objects in 5D EGB gravity which to our knowledge do not appear in the literature. We do not pursue this aspect now however we indicate a work-around the presence of the undesirable central singularity.  \\
    \item[(iv)] The physical admissibility of our solutions require that the following energy conditions, viz., (i)weak energy condition (WEC), (ii) strong energy
condition (SEC) and (iii) dominant energy condition (DEC) hold true at each interior point of the star.
These conditions are equivalent to
\begin{eqnarray}
\text{WEC}: \rho - p \geq 0,~~~\text{SEC}: \rho+p \geq  0,~~~
\text{DEC}: \rho+4p \geq  0.\label{eq24}
\end{eqnarray}
It is clear from Fig. \ref{f4}. that all three energy conditions are satisfied for the 5-$d$ model we have developed and for our choice of parametric space.\\
\item[(v)] In the absence of anisotropy or dissipation it is well-known that there has no upper mass limit if the EoS has an adiabatic index $\Gamma > 4/3$ where

\begin{equation}
\Gamma=\frac{p+\rho}{p}\,\frac{dp}{d\rho}\, ,\label{eq31}
\end{equation}
characterising stable regions within the fluid sphere. This result has been extended to include pressure anisotropy, viscosity and heat flow by Chan and co-workers \cite{st1,st2}. They showed that dissipation in particular, renders the system less unstable by diminishing the total mass entrapped inside the fluid sphere. Fig. \ref{f5} reveals an interesting peculiarity in the 5D stability index. The central portion of the star appears to be more stable than regions greater than some finite radius away from the center. This behaviour is unphysical in a realistic stellar model as the central density is highest here and would expect more heating and thus thermodynamically unstable regions. As one moves away to the cooler regions, heat generation would drop-off thus leading to stable regions. The adiabatic index points to the fact that our solutions would be suitable in describing a core-envelope model of a compact object in EGB gravity. \\
\item[(vi)] The mass profile for the 5D model is plotted in Fig. \ref{f6}. We observe that the mass function increases as one tends away from the centre of the stellar configuration with $m(0) = 0$ which is expected as more mass is contained within larger concentric shells. \\
\item[(vii)] A star remains in static equilibrium under the forces namely, gravitational force and  hydrostatics force,  if the summation of two active forces on the system is zero. Along these directions, a wide class of astrophysical solutions 
(including wormholes/compact stars) have been studied  ( see, e.g., 
Refs. \cite{Rahaman:2013xoa,Rahaman:2014dpa,Sarkar:2019lay,Maurya:2018kxg}).
This condition is formulated mathematically via,
\begin{eqnarray}\label{mtov}
-\frac{dp}{dx}-{\dot{y}\over y}(\rho+p)=0, ~~~\text{or}~~~F_h+F_g =0, 
\end{eqnarray}
where the first term represents the hydrodynamic force ($F_h $) and the second term corresponds to gravitational force ($F_g$), respectively. The variation of the forces is shown in Fig.  \ref{f9} and our system is stable in terms of the equilibrium of forces.
\end{enumerate}

The analysis above suggests that the conformally flat stellar model we have derived does exhibit pleasing physical behaviour consistent with the elementary  requirements. In contrast with the other conformally flat solution the Schwarzschild interior, this solution displays a variable energy density and consequently does not suffer from the pathology of an infinite sound speed. 

\begin{figure}[h]
\centering
\includegraphics[width=1\linewidth]{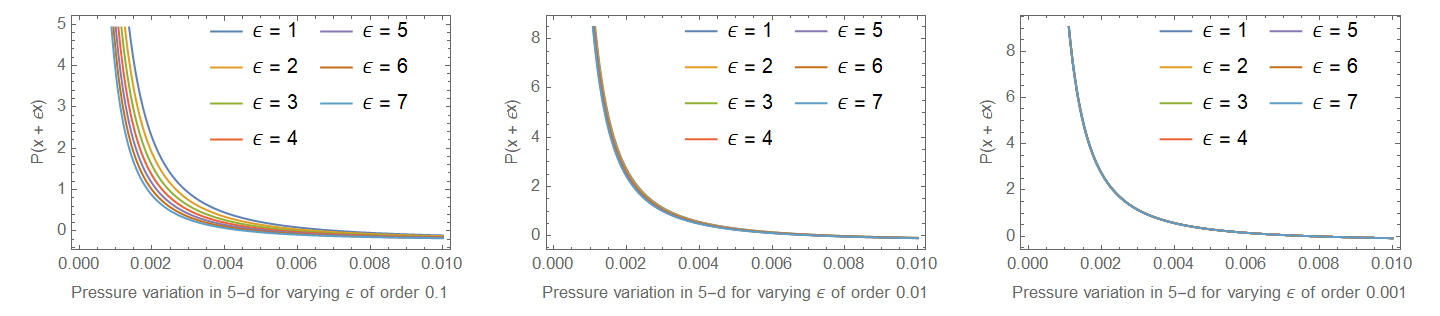}\\
\includegraphics[width=1\linewidth]{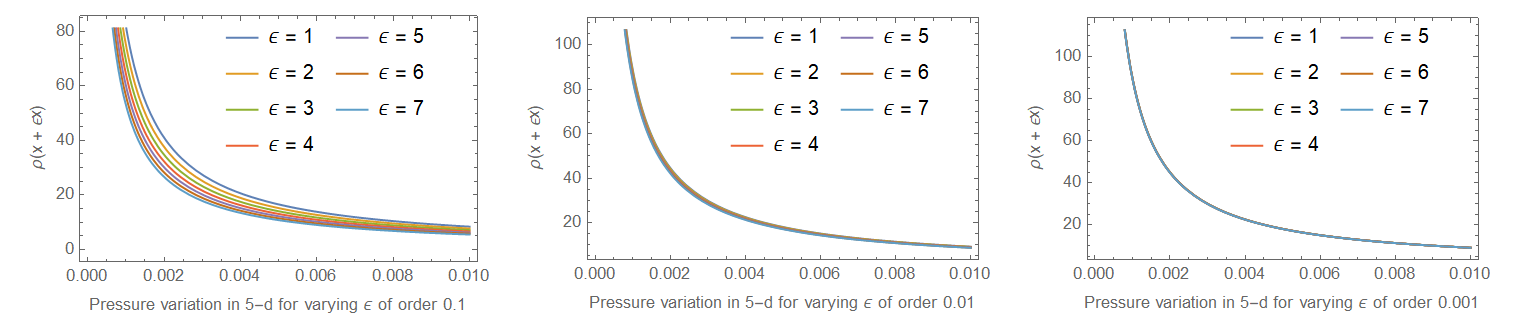}\\
\includegraphics[width=1\linewidth]{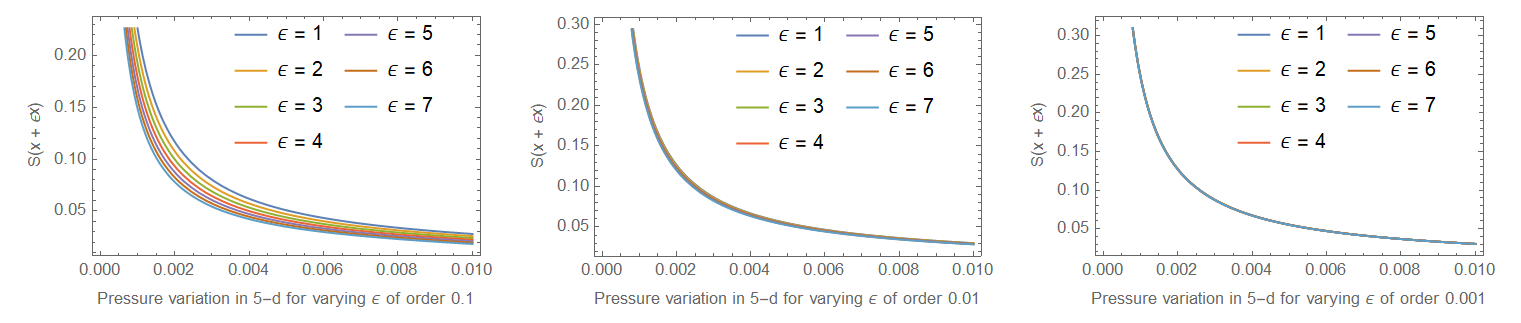}
\caption{Pressure, energy density and speed of sound versus radial pulsation $x + \epsilon x$ for $\epsilon$ of the order  $0.1$, $ 0.01$ and $0.001$, respectively. The labels of the curves are the same as given in Fig. \ref{f1}.} \label{f7}
\end{figure}

\section{Six dimensional case}\label{sec6}

 We now turn our attention to the remaining critical dimension in EGB namely the six dimensional case. The EGB field equations for a six dimensional metric
 \begin{equation}
ds^{2} = -e^{2 \nu} dt^{2} + e^{2 \lambda} dr^{2} + r^{2}d\Omega^2, \label{5}
\end{equation}
where $d\Omega^2= d\theta^{2} + \sin^{2} \theta d \phi^{2} + \sin^{2} \theta \sin^{2} \phi d\psi^{2} + \sin^{2} \theta \sin^{2} \phi \sin^{2}\psi d\eta^{2}$
 (\ref{2}) when expanded amount to the system
\begin{eqnarray}
\rho &=& \frac{1}{e^{4 \lambda} r^{4}} \left[ ( 4 r^{3} e^{2 \lambda} - 48 \alpha r ( 1 - r e^{2 \lambda}) ) \lambda^{\prime} - 6 r^{2} e^{2 \lambda} (1 - e^{2 \lambda}) + 12 \alpha (e^{2 \lambda} - 1)^{2} \right],  \label{7a} \\ \nonumber \\
p &=&  \frac{1}{e^{4 \lambda} r^{4}} \left[ (1 - e^{2 \lambda}) (6 r^{2} e^{2 \lambda} - 48 \alpha r \nu^{\prime} + 12 \alpha e^{2 \lambda} - 12 \alpha) + 4 r^{3} e^{2 \lambda} \nu^{\prime} \right],  \label{7b} \\ \nonumber \\
p &=& \frac{1}{e^{4 \lambda} r^{2}} \left( (12 \alpha (e^{2 \lambda} - 1) + r^{2} e^{2 \lambda}) (\nu^{\prime \prime} + (\nu^{\prime})^{2} - \nu^{\prime} \lambda^{\prime}) + 24 \alpha \nu^{\prime} \lambda^{\prime} \right) \nonumber \\
                 & \quad & + \frac{1}{e^{4 \lambda} r^{3}} \left( (3 r^{2} e^{2 \lambda} + 12 \alpha (e^{2 \lambda} - 1)) (\nu^{\prime} - \lambda^{\prime}) + 3 r e^{2 \lambda} (1 - e^{2 \lambda}) \right), \label{7c}
\end{eqnarray}
in the canonical spherical coordinates $ (x^{a}) $. The nonlinearity in the system (\ref{7a})--(\ref{7c}) has now greatly increased because of the presence of the EGB coupling parameter $ \alpha $ and further dynamical terms that were suppressed in 5-d. 

The system (\ref{7a})--(\ref{7c}) may be converted to the equivalent form 
\begin{eqnarray}
\frac{ 12\beta x (Z - 1) \dot{Z} - 4 x^{2} \dot{Z} - 6 x (Z - 1) + 3\beta (Z - 1)^{2} }{x^{2}} &=& \frac{\rho}{C}, \label{9a} \\ \nonumber \\
 \frac{\left( 24\beta x (1 - Z) + 8 x^{2} \right) Z \dot{y} + (Z - 1) (6x + 3\beta (1 - Z)) y}{x^{2} y}  &=& \frac{p}{C}, \label{9b} \\ \nonumber \\
 4 x^{2} Z \left( x + 3\beta [1-Z] \right) \ddot{y} + 2 x \left( x^{2} \dot{Z} + 3 \beta \left[  (1 - 3 Z) \dot{Z}x - 2 Z ( 1 - Z) \right] \right) \dot{y} \nonumber \\ + 3 \left(\beta (1-Z)+x\right)\left(\dot{Z}x-Z+1\right)  y &=& 0.  \label{9d}
\end{eqnarray}
 through the coordinate change $ x = C r^{2} $, $ e^{2 \nu} = y^{2} (x) $ and $ e^{- 2 \lambda} = Z(x) $. Note that  (\ref{9d}) is the equation of pressure isotropy in six-dimensional EGB theory and we have redefined $\beta = 4\alpha C$. It has been expressed as a linear second order differential equation in $ y $ (if $ Z $ is a known quantity) which is the great advantage of this coordinate choice.  The condition of pressure of isotropy may also be expressed as 
\begin{eqnarray}
\left[ x^{2} (2 x \dot{y} + 3 y) + 3\beta x (2 x \dot{y} + y - (6 x \dot{y} + y) Z) \right] \dot{Z} \nonumber \\
- 3 \beta \left[ 4 x^{2} \ddot{y} - 4 x \dot{y} - y \right] Z^{2} + \left[ x (4 x^{2} \ddot{y} - 3 y) + 6\beta (2 x^{2} \ddot{y} - 2 x \dot{y} - y) \right] Z 
+  3 (x + \beta) y &=& 0 \label{9e}
\end{eqnarray}
 a nonlinear first order differential in $ Z $ (if $y$ is a known quantity). We observe that equation (\ref{9a}) is easily integrated if we assume that the density is constant, ie., $\rho = \rho_0$. Integration of (\ref{9a}) yields
 \begin{eqnarray}
 Z(x) &=& 1 + \frac{x}{3\beta} \left(1 \pm \sqrt{\frac{9c_1\beta^2 }{x^{\frac{5}{2}}} + \frac{3\beta \rho_0 + 5C}{5C} }  \right) , \label{102}
 \end{eqnarray} 
 where $c_1$ is a constant of integration. Note that (\ref{102}) when written in the form 
 \be
 Z = 1 + \frac{x}{3\beta} \pm \frac{1}{3\beta} \sqrt{\frac{9c_1 \beta^2}{x^{1/2}}+ \left(1 + \frac{3\beta \rho_0}{5C}  \right)x^2}
 \ee 
 reveals a singularity at the centre $x = 0$ unlike the 5-dimensional case which was singularity-free. This point further illustrates the manifest different physics inherent in 5 and 6 dimensional EGB gravity. Note that the six dimensional vacuum solution of Boulware and Deser \cite{boul} is regained from result (\ref{102}) by setting the constant density $\rho_0 = 0$ and is given by 
  \begin{eqnarray}
 Z(x) &=& 1 + \frac{x}{3\beta} \left(1 \pm \sqrt{1 + \frac{ 9c_1\beta^2 }{x^{\frac{5}{2}}}  }  \right).  \label{102a}
 \end{eqnarray} 
 Bogdanos {\it{et al}} \cite{bogd} have analysed the 6--dimensional case in EGB and demonstrated the validity of Birkhoff's theorem for this order.
 
 Equation (\ref{9e}) is an Abel differential equation of the second kind and admits only a few exact solutions for certain cases. 
Inserting (\ref{111}) into the pressure isotropy equation gives 
\begin{equation}
\left(x \dot{Z}-Z+1\right) \left(-3 \beta \sqrt{Z} \sinh f(x) +(3 \beta +2 x) \cosh f(x) -6 \beta Z \cosh f(x) \right) = 0 \label{150}
\end{equation}
after factorization. As for the 5 dimensional case, the Schwarzschild interior solution emanates from the vanishing of the first factor on the left hand side of (\ref{150}). The other factor generates the condition 
\begin{equation}
\tanh f(x) = \frac{2x + 3\beta (1 -2Z)}{3\beta \sqrt{Z}} \label{151}
\end{equation}
which has a remarkably similar form similar to that  for the 5 dimensional case despite the presence of extra terms in the pressure isotropy equation. 
Following the same procedures as in the previous section the general solution of (\ref{151}) has the form 
\begin{equation}
Z =\frac{(e^{4 x} + h^2) (15 \beta +8 x)-2h (9 \beta +8 x)  e^{2 x }}{24 \beta  \left(e^{2 x}-h\right){}^2} 
\pm  \frac{ \sqrt{ \left(e^{2 x}+h \right){}^2 \left((e^{4 x} +h^2) (27 \beta +16 x)-2 h(21 \beta +16 x) e^{2 x}\right)}}{8 \sqrt{ 3\beta}  \left(e^{2 x}-h\right){}^2} \label{152}
\end{equation}
where $h$ is a constant of integration. Once again the form of the temporal potential may be obtained via equation (\ref{111}) albeit in terms of an unresolved integral. 

We calculate the physical quantities relevant to our investigation. Not unexpected, these expressions are unwieldy on account of the long expressions for the gravitational potentials. The energy  density is given by   
\begin{eqnarray}
\frac{\rho}{C} &=& \left[4 L_0 x \left(-9 k^3 \left(\sqrt{L_3 \left(k+e^{2 x}\right)^2}-\sqrt{\beta } k^2\right)+k^2 e^{2 x} \left(\sqrt{\beta } k^2 (40 \beta +72 x-9)+(27-16 \beta ) \sqrt{L_3 \left(k+e^{2 x}\right)^2}\right)  \right. \right.   \nonumber\\
&& \left. \left.  
-k e^{4 x} \left(6 \sqrt{\beta } k^2 (-4 \beta +12 x+3)+(16 \beta +27) \sqrt{L_3 \left(k+e^{2 x}\right)^2}\right)+3 e^{6 x} \left(2 \sqrt{\beta } k^2 (4 \beta -12 x+3) \right.  \right. \right. \nonumber\\
&& \left. \left. \left.  
+3 \sqrt{L_3 \left(k+e^{2 x}\right)^2}\right)+\sqrt{\beta } k e^{8 x} (40 \beta +72 x+9)-9 \sqrt{\beta } e^{10 x}\right)\right] \times \frac{1}{x^2}+\frac{\left(-L_0\right){}^2}{27 \beta  x^2 \left(e^{2 x}-k\right)^4}     \nonumber\\
&& 
+\frac{2 x \left(4 \beta  \left(k^2-4 k e^{2 x}+e^{4 x}\right)-9 x \left(e^{2 x}-k\right)^2+2 \sqrt{\beta } L_1\right)}{3 \beta  x^2 \left(e^{2 x}-k\right)^2}
\Big/ \left[\frac{27 \beta  \left(e^{2 x}-k\right)^5 \sqrt{L_3 \left(k+e^{2 x}\right)^2}}{x^2}  \right.   \nonumber\\
&& \left. 
-4 \left(\frac{\frac{9 \left(k^2-2 k e^{2 x} (2 x+1)+e^{4 x} (4 x+1)\right)}{\beta }+\frac{L_5}{\sqrt{\beta } \sqrt{L_3 \left(k+e^{2 x}\right)^2}}-4 k e^{2 x}+20 e^{4 x}}{9 \left(e^{2 x}-k\right)^2}-\frac{4 L_0 e^{2 x}}{9 \beta  \left(e^{2 x}-k\right)^3}\right)\right],
\end{eqnarray}
while the isotropic particle pressure assumes the form 
\begin{eqnarray}
\frac{p}{C} &=& \frac{L_0 \left(4 \beta  k^2+9 k^2 x+2 \sqrt{\beta } \sqrt{L_3 \left(k+e^{2 x}\right)^2}-2 k e^{2 x} (8 \beta +9 x)+e^{4 x} (4 \beta +9 x)\right)}{27 \beta  x^2 \left(e^{2 x}-k\right)^4}  \nonumber\\
&&
+ \frac{1}{27 x^2 \left(e^{2 x}-k\right)^2}
 \left[8 \sqrt{\frac{9 \beta  \left(k^2-2 k e^{2 x}+e^{4 x}\right)+L_0}{\beta  \left(e^{2 x}-k\right)^2}} \left(2 \beta  k^2-3 k^2 x+\sqrt{\beta } \sqrt{L_3 \left(k+e^{2 x}\right)^2} \right.  \right.  \nonumber\\
&& \left.  \left. 
+2 k e^{2 x} (3 x-4 \beta )+e^{4 x} (2 \beta -3 x)\right) \left(\sqrt{\frac{9 \beta  \left(k^2-2 k e^{2 x}+e^{4 x}\right)+L_0}{\beta  \left(e^{2 x}-k\right)^2}}+L_4\right)\right].
\end{eqnarray}

%%%%%%%%%%%%%%%%%%%%%%%%%%%%%%%%%%%%%%%%%%%%%%%%%%%%%%%%%%%%%%%%%
The expressions below will assist us analyse the energy conditions. 
\begin{eqnarray}
\frac{\rho-p}{C} &=& \left[2  \left(12 \sqrt{\beta } k^6 L_1 \left(-4 \beta ^2+81 x^2+17 \beta  x\right)-2 k^5 e^{2 x} \left(k^2 \left(270 \beta  (8 x+5) x^2-3645 x^3+48 \beta ^2 (23 x+32) x \right.\right. \right.\right. \nonumber\\
&& \left.\left.\left.\left.
+16 \beta ^3 (4 x+9)\right)-8 \sqrt{\beta } L_1 \left(27 (2 x-9) x^2+\beta ^2 (33-4 x)+6 \beta  (5 x+6) x\right)\right)  \right.\right.  \nonumber\\
&& \left.\left.
+2 k^4 e^{4 x} \left(3 k^2 \left(720 \beta  (2 x+5) x^2-2835 x^3+8 \beta ^2 (199-44 x) x-32 \beta ^3 (10 x+1)\right)  \right.\right. \right. \nonumber\\
&& \left.\left. \left.
+2 \sqrt{\beta } L_1 \left(27 (45-32 x) x^2+4 \beta ^2 (104 x-33)+9 \beta  (64 x-49) x\right)\right)+6 k^3 e^{6 x} \left(k^2 \left(90 \beta  (8 x-57) x^2+2835 x^3  \right.\right. \right. \right.  \nonumber\\
&& \left.\left. \left. \left. +16 \beta ^2 (23 x-84) x+16 \beta ^3 (1-20 x)\right)+32 \sqrt{\beta } L_1 x \left(-6 \beta ^2+27 x^2-29 \beta  x\right)\right)
\right.\right.    \nonumber\\
&& \left.\left. 
-4 \sqrt{\beta } k^2 e^{8 x} \left(32 \sqrt{\beta } k^2 x (9 x-\beta ) (15 x-2 \beta )+L_1 \left(27 (32 x+45) x^2-4 \beta ^2 (104 x+33)-9 \beta  (64 x+49) x\right)\right) \right.\right.   \nonumber\\
&& \left.\left. 
+2 k e^{10 x} \left(3 k^2 \left(90 \beta  (8 x+57) x^2-2835 x^3+16 \beta ^2 (23 x+84) x-16 \beta ^3 (20 x+1)\right)  \right.\right. \right.  \nonumber\\
&& \left.\left. \left.
+8 \sqrt{\beta } L_1 \left(27 (2 x+9) x^2-\beta ^2 (4 x+33)+6 \beta  (5 x-6) x\right)\right)-6 e^{12 x} \left(k^2 \left(-32 \beta ^3-45 (32 \beta +63) x^3  \right.\right. \right. \right. \nonumber\\
&& \left.\left. \left. \left.
+16 \beta  (22 \beta +225) x^2+8 \beta ^2 (40 \beta +199) x\right)+2 \sqrt{\beta } L_1 \left(-4 \beta ^2+81 x^2+17 \beta  x\right)\right)
\right.\right.   \nonumber\\
&& \left.\left. 
-3 k^8 (4 \beta +9 x) \left(8 \beta ^2+45 x^2+20 \beta  x\right)-2 k e^{14 x} \left(270 \beta  (8 x-5) x^2+3645 x^3+48 \beta ^2 (23 x-32) x  \right.\right.  \right. \nonumber\\
&& \left.\left.\left.
+16 \beta ^3 (4 x-9)\right)+3 e^{16 x} (4 \beta +9 x) \left(8 \beta ^2+45 x^2+20 \beta  x\right)\right)\right]  \nonumber  \\ 
&& 
\Big/  \frac{9 x^2 \left(e^{2 x}-k\right)^2}{243 \beta  L_3 x^4 \left(e^{2 x}-k\right)^7 \left(k+e^{2 x}\right) -8 c L_4 \sqrt{\frac{L_3-2 \sqrt{\beta } L_1}{\beta  \left(e^{2 x}-k\right)^2}} \left(L_2 +\sqrt{\beta } L_1\right)},
\end{eqnarray}

\begin{eqnarray}
\frac{\rho+p}{C} &=& \left[8  \left(-k^6 L_1 \left(4 \beta ^2+27 x^2+51 \beta  x\right)+4 k^5 e^{2 x} \left(L_1 \left(27 (2 x+1) x^2-\beta ^2 (4 x+13)+6 \beta  (5 x+3) x\right) \right.\right. \right. \nonumber\\
&& \left.\left.\left.
-\sqrt{\beta } k^2 \left(27 (10 x+7) x^2+2 \beta ^2 (4 x-27)+6 \beta  (23 x-11) x\right)\right)+k^4 e^{4 x} \left(L_1 \left(-27 (32 x+5) x^2 \right.\right. \right. \right. \nonumber\\
&& \left.\left. \left.\left.
+52 \beta ^2 (8 x+1)+9 \beta  (64 x+1) x\right)-4 \sqrt{\beta } k^2 \left(-108 (5 x+2) x^2+8 \beta ^2 (15 x+11)+3 \beta  (44 x+85) x\right)\right) \right. \right.  \nonumber\\
&& \left.\left. 
+4 k^3 e^{6 x} \left(\sqrt{\beta } k^2 \left(27 (10 x-3) x^2+2 \beta ^2 (43-60 x)+6 \beta  (23 x+39) x\right)+12 L_1 x \left(-6 \beta ^2+27 x^2-29 \beta  x\right)\right) \right.\right.   \nonumber\\
&& \left.\left. 
-k^2 e^{8 x} \left(32 \sqrt{\beta } k^2 x (9 x-\beta ) (15 x-2 \beta )+L_1 \left(27 (32 x-5) x^2+52 \beta ^2 (1-8 x)+9 \beta  (1-64 x) x\right)\right) \right.\right.   \nonumber\\
&& \left.\left. 
+4 k e^{10 x} \left(\sqrt{\beta } k^2 \left(27 (10 x+3) x^2-2 \beta ^2 (60 x+43)+6 \beta  (23 x-39) x\right)+L_1 \left(27 (2 x-1) x^2  \right.\right. \right.\right. \nonumber\\
&& \left.\left. \left.\left.
+\beta ^2 (13-4 x)+6 \beta  (5 x-3) x\right)\right)+e^{12 x} \left(4 \sqrt{\beta } k^2 \left(108 (5 x-2) x^2+8 \beta ^2 (11-15 x)+3 \beta  (85-44 x) x\right) 
\right.\right. \right. \nonumber\\
&& \left.\left. \left.
+L_1 \left(4 \beta ^2+27 x^2+51 \beta  x\right)\right)-2 \sqrt{\beta } k^8 (\beta -12 x) (4 \beta +9 x)-4 \sqrt{\beta } k e^{14 x} \left(27 (10 x-7) x^2
\right.\right. \right. \nonumber\\
&& \left.\left. \left.
+2 \beta ^2 (4 x+27)+6 \beta  (23 x+11) x\right)+2 \sqrt{\beta } e^{16 x} (\beta -12 x) (4 \beta +9 x)\right)\right]
 \nonumber \\  &&  \Big/
\frac{8  L_4 \left(\sqrt{\beta } L_1+L_2\right) \sqrt{\frac{L_3-2 \sqrt{\beta } L_1}{\beta  \left(e^{2 x}-k\right)^2}}}{9 x^2 \left(e^{2 x}-k\right)^2}+27 \sqrt{\beta } x^2 \left(e^{2 x}-k\right)^5 L_5,
\end{eqnarray}

\begin{eqnarray}
\frac{\rho+5p}{C} &=&\frac{40 L_4 \left(\sqrt{\beta } L_1+L_2\right) \sqrt{\frac{L_3-2 \sqrt{\beta } L_1}{\beta  \left(e^{2 x}-k\right)^2}}}{9 x^2 \left(e^{2 x}-k\right)^2}
-\left[4 \left(6 \sqrt{\beta } k^6 L_1 \left(-4 \beta ^2+189 x^2+85 \beta  x\right) +2 k^5 e^{2 x} \left(k^2 \left(54 \beta  (20 x+17) x^2
\right. \right. \right. \right. \nonumber\\
&& \left. \left. \left. \left.
+3645 x^3+24 \beta ^2 (23 x-97) x+8 \beta ^3 (4 x-99)\right)+4 \sqrt{\beta } L_1 \left(-27 (2 x+21) x^2+\beta ^2 (4 x+105)+6 \beta  (3-5 x) x\right)\right)
\right. \right.   \nonumber\\
&& \left. \left. 
-2 k^4 e^{4 x} \left(3 k^2 \left(144 \beta  (5 x-19) x^2+2835 x^3-4 \beta ^2 (44 x+653) x-160 \beta ^3 (x+2)\right)
\right. \right. \right.  \nonumber\\
&& \left. \left. \left. 
+\sqrt{\beta } L_1 \left(-27 (32 x+105) x^2+4 \beta ^2 (104 x+105)+9 \beta  (64 x+101) x\right)\right)-6 k^3 e^{6 x} \left(k^2 \left(328 \beta ^3+45 (8 \beta -63) x^3
\right. \right. \right. \right. \nonumber\\
&& \left. \left. \left. \left.
+2 \beta  (92 \beta +2403) x^2+40 \beta ^2 (57-4 \beta ) x\right)+16 \sqrt{\beta } L_1 x \left(-6 \beta ^2+27 x^2-29 \beta  x\right)\right)
\right. \right.   \nonumber\\
&& \left. \left. 
+2 \sqrt{\beta } k^2 e^{8 x} \left(32 \sqrt{\beta } k^2 x (9 x-\beta ) (15 x-2 \beta )+L_1 \left(420 \beta ^2-9 (64 \beta +315) x^2+864 x^3+\beta  (909-416 \beta ) x\right)\right)
\right. \right.   \nonumber\\
&& \left. \left.  
-2 k e^{10 x} \left(3 k^2 \left(-328 \beta ^3+45 (8 \beta +63) x^3+2 \beta  (92 \beta -2403) x^2-40 \beta ^2 (4 \beta +57) x\right)
\right. \right. \right.  \nonumber\\
&& \left. \left. \left. 
+4 \sqrt{\beta } L_1 \left(27 (2 x-21) x^2+\beta ^2 (105-4 x)+6 \beta  (5 x+3) x\right)\right)-6 e^{12 x} \left(k^2 \left(144 \beta  (5 x+19) x^2-2835 x^3
\right. \right. \right. \right. \nonumber\\
&& \left. \left. \left. \left. 
+4 \beta ^2 (653-44 x) x-160 \beta ^3 (x-2)\right)+\sqrt{\beta } L_1 \left(-4 \beta ^2+189 x^2+85 \beta  x\right)\right)-3 k^8 (4 \beta +9 x) 
\right. \right.   \nonumber\\
&& \left. \left. 
\times \left(4 \beta ^2+45 x^2+68 \beta  x\right)+2 k e^{14 x} \left(792 \beta ^3+135 (8 \beta -27) x^3+6 \beta  (92 \beta -153) x^2+8 \beta ^2 (4 \beta +291) x\right)  \right. \right.   \nonumber\\
&& \left. \left.  
+3 e^{16 x} (4 \beta +9 x) \left(4 \beta ^2+45 x^2+68 \beta  x\right)\right) \times 27 \beta  L_3 x^2 \left(e^{2 x}-k\right)^5 \left(k+e^{2 x}\right)\right],
\end{eqnarray}
For simplicity of notations  we have used \\
$L_1^2= k^4 (4 \beta +9 x)+4 \beta  k^3 e^{2 x}-18 k^2 e^{4 x} x+4 \beta  k e^{6 x}+e^{8 x} (4 \beta +9 x),$\\
$L_2= k^2 (2 \beta -3 x)+2 k e^{2 x} (3 x-4 \beta )+e^{4 x} (2 \beta -3 x),$\\
$L_3=k^2 (4 \beta +9 x)-2 k e^{2 x} (2 \beta +9 x)+e^{4 x} (4 \beta +9 x)$\\
$L_5= k^3 (4 \beta +9 x)-9 k^2 e^{2 x} x-9 k e^{4 x} x+e^{6 x} (4 \beta +9 x),$\\
$L_4=\tanh \left(B+\frac{1}{2} \int_0^x \frac{3\beta (e^{2x}-k)}{x \sqrt{L_3-2 (e^{2x}+k) \sqrt{\beta L_3 }} } \, dx\right).$

We have elected not to include the expressions for the adiabatic stability parameter $\Gamma$, mass profile as well as the sound-speed squared as these expressions are extremely long. Nevertheless, we have plotted the behaviour of these quantities and have studied them carefully in what follows. 

%%%%%%%%%%%%%%%%%%%%%%%%%%%%%%%%%%%%%%%%%%%%%%%%%%%%%%%%%%%%%%%%%%%%%%
\begin{figure}[h]
\centering
\includegraphics[width=1\linewidth]{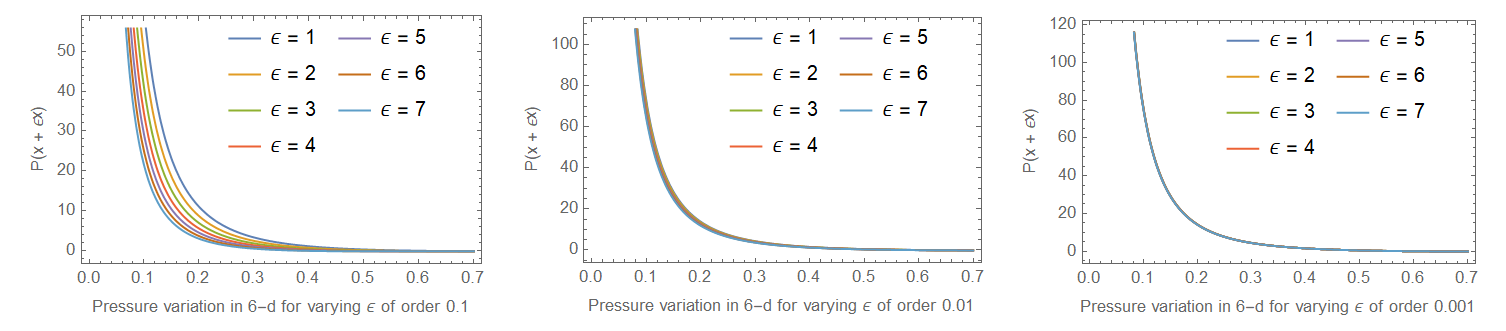}\\
\includegraphics[width=1\linewidth]{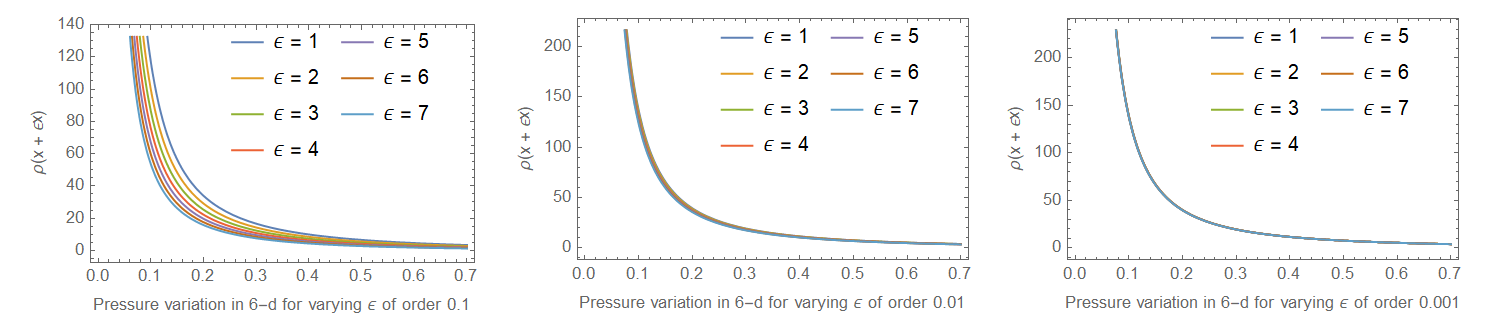}\\
\includegraphics[width=1\linewidth]{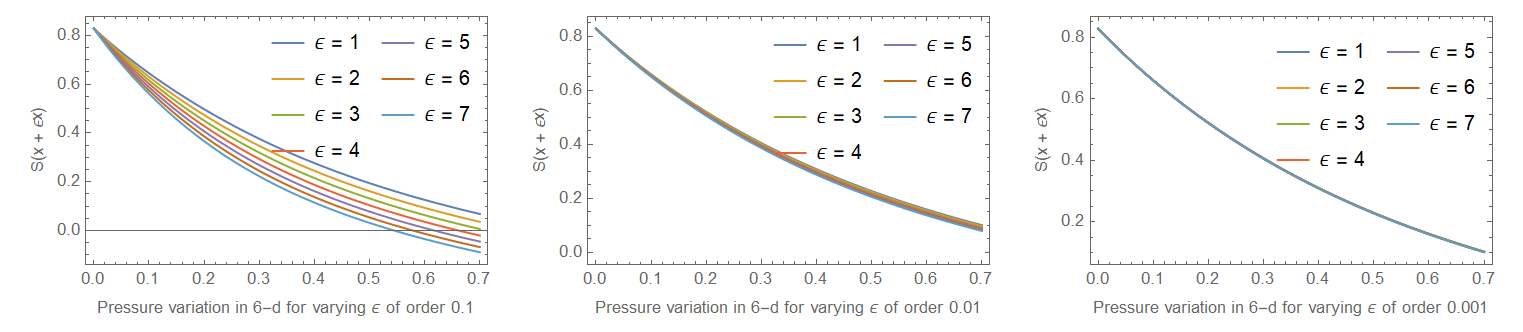}
\caption{Pressure, energy density and speed of sound versus radial pulsation $x + \epsilon x$ for $\epsilon$ of the order  $0.1$, $0.01$ and $0.001$, respectively. The labels of the curves are the same as given in Fig. \ref{f1}.} \label{f8}
\end{figure}
\subsection{Physical analysis of the 6-$d$ hypersphere}

 For the purposes of graphical plots in 6-d the following parameter values were obtained for pleasing physical behaviour:   $A=2, ~B=1,~C=0.1,~k= -0.9$ and $\beta= 8$. 
Observations of the right panels of each of the figures from 1 to 6 reveal interesting properties of the 6-$d$ hypersphere.
\begin{enumerate}
    \item [(i)]The boundedness of the solution is revealed in Fig. \ref{f1} which illustrates that the pressure decreases monotonically from the centre of the star and vanishes at the boundary. This qualifies the model as a potential compact star model suitable for applicability to neutron stars or cold planets.  \\
    \item[(ii)] The density  plotted in Fig. \ref{f2} increases sharply at the centre of the configuration and drops off smoothly towards the surface of the stellar surface.  This   behaviour is not inconsistent with stellar structure dynamics as the interior processes are not known. Importantly the density remains positive definite within the distribution.  \\
    \item[(iii)] The causality criterion $0 \leq v_s = \frac{dp}{d\rho} \leq 1$ holds everywhere inside the fluid sphere as observed in Fig. \ref{f3}. We observe that the sound speed decreases from the centre outwards attaining a minimum value at the surface of the star. This could be due to phase transitions leading to different equations of state in different regions of the star \cite{satya1}. \\
    \item[(iv)] The  (i)weak energy condition (WEC), (ii) strong energy
condition (SEC) and (iii) dominant energy condition (DEC) given by 
\begin{eqnarray}
\text{WEC}: \rho - p \geq 0,~~~\text{SEC}: \rho+p \geq  0,~~~
\text{DEC}: \rho+5p \geq  0.\label{eq24}
\end{eqnarray} should hold everywhere inside the star. We observe that all energy conditions are satisfied throughout the interior of the star for the 6-$d$ hypersphere. \\
\item[(v)] The stability index plotted in Fig. \ref{f5} shows that the surface layers are more stable than the central regions. This is expected as the density is highest at the center making energy-generating processes such as nuclear fusion more efficient leading to more heat generation here. The heated core is likely to be more unstable than the surface of the star. The requirement $\Gamma > \frac{4}{3}$ is always satisfied as depicted in Fig. \ref{f5}.   \\
\item[(vi)] From Fig. \ref{f6} it may be  observed that the mass function increases as one moves away from the centre of the stellar configuration with $m(0) = 0$ which is expected as more mass is contained within larger concentric shells.  \\
\item[(vii)] Using the expression (\ref{mtov}) the modified form of the TOV equation has been plotted in Fig.  \ref{f9}.
We observe that the system is completely stable under the equilibrium of the different forces, i.e., $F_h+F_g =0$, with respect to the radial coordinate $x$.
\end{enumerate}

The plots above demonstrate the viability of the new 6-d conformally flat solution in representing realistic stellar distributions. The persistent central singularity may be eliminated by the inclusion of a suitable nonsingular core and its attendant matching requirements between the different layers. 

\subsection{Stability under radial pulsations}

We combine our comments on the question of whether the models developed are stable under radial pulsations.  Historically,  stability was analysed extensively   by Chandrasekhar \cite{chandra} who developed a test for stability under perturbations of the pressure and energy density for a  linearised system of Einstein's equations. This scheme is only applicable to the Einstein's equations and not directly useful in EGB theory. It would be useful  to carry out the computations of Chandrasekhar within higher curvature gravity.  Bardeen \textit{et al}  devised a catalogue of methods to study the normal modes of radial pulsations of stellar models. A normal mode of radial pulsation for an equilibrium solution has the form $\delta r = \xi (r) \exp (i \sigma t)$ where $\xi$ is a trial function and the eigenvalue $\sigma$ is the frequency.   Additionally it is assumed, for this process that, the Lagrangian change of the pressure vanishes at the boundary and that the trial function is finite at the centre.  

It is worth noting that stability can take on different meanings depending on the context. Ample  warnings are contained in the work of Bardeen \textit{et al} \cite{bardeen}.  Kokkotas and Ruoff \cite{kokkotas} make use of a numerical technique and give two approaches to studying stability: the shooting method and the method of finite differences.    In the context of EGB gravity the stability of scalarized black holes has been considered by Silva \textit{et al} \cite{silva}.   Since there is  equivalent criteria of the Chandrasekhar integral condition for EGB gravity, we shall analyse our exact solution under radial pulsations by perturbing the radial coordinate $x$ as $x + \epsilon x$, where $\epsilon$ is a very small quantity in relation to  the stellar radius, in the energy density, pressure and sound speed. It will be immediately apparent that for small values of $\epsilon$ the dynamical quantities converge to the original shapes.  We provide plots where $\epsilon$ is of the order $0.1, 0.01, 0.001$ in order to confirm that our solution is indeed stable under radial pulsations. This follows the  same spirit as the well known Lyapunov stability of curves and mimics the approach of Herrera \cite{st2,herrera} in discussing the concept of cracking in anisotropic spheres. 

Fig. \ref{f7} and \ref{f8} depicts the energy density, pressure and sound speed  profiles for the 5 and 6 dimensional conformally flat  EGB models  under pulsations of the type $x \rightarrow x + \epsilon x$ for three different orders of magnitude of  $\epsilon$  namely $0.1, 0.01, 0.001$ and each plot showing increments up to seven curves. In other words, for each profile we have plotted seven increments to check the trend:  for example, for the order of magnitude 0.001 we show the curves for $\epsilon = 0.001, 0.002, 0.003, 0.004, 0.005, 0.006, 0.007$.   It is clear that in all cases as $\epsilon$ decreases the perturbed curves converge  toward the equilibrium curve $\epsilon = 0$. This is the case also for larger orders of magnitude $\epsilon = 0.01$ and $0.1$.  This may be understood to mean that  these dynamical quantities are stable under radial pulsations. It must be borne in mind that we have achieved an exact solution in closed form only for $Z$ whereas the form for $y$ is expressed as an integral. Consequently, we had to resort to numerical techniques thus necessitating the check whether small disturbances of the system generate severe deviations from the equilibrium position.

\section{Conclusion}\label{sec7}

We have thoroughly investigated conformally flat static isotropic spacetimes in Einstein--Gauss--Bonnet spacetimes.  Our conclusions may be summarised as follows: 
\begin{description}
\item{(i)} The interior  Schwarzschild metric generates a constant density hypersphere in EGB, however,  the hypothesis of a constant density results in  a generalized Schwarzschild  solution.  In the 5 dimensional case the general solution is completely free of singularities at the stellar centre but a coordinate central  singularity in the 6 dimensional case is unavoidable.  
\item{(ii)}  The assumption of constant density leads to the generalised Schwarzschild solutions for 5 and 6 dimensions as mentioned in (i). It has been verified by direct substitution that the generalised Schwarzschild solutions for 5 and 6 dimensions result in the conformal flatness equation being identically satisfied. So the general constant density metric is indeed conformally flat. Conversely, assuming conformal flatness generates two branches of solutions  one of which is Schwarzschild and the other is a new solution which does not yield constant density. The temporal gravitational potential is expressed as an integral, however, with the aid of numerical techniques it is possible to analyse the physical properties of the solution graphically. Suitable parameter spaces are exhibited demonstrating that the models in 5 and 6 dimensions satisfy basic physical requirements demanded of stellar distributions.   
\item{(iii)} Finally,  the Schwarzschild metric is known to be conformally flat. However, the assumption of conformal flatness leads to two branches of solutions only one of which is Schwarzschild. As mentioned in (ii) above, the non-Schwarzschild solution is shown to be reasonably behaved in both 5 and 6 dimensions. 
\end{description} 
When subjected to an elementary test of stability with respect to radial pulsations, both five and six dimensional models converged to the equilibrium position verifying its stability.  The new conformally flat solution did not exhibit the defect of an infinite sound speed as its counterpart the Schwarzschild metric does in standard Einstein gravity. Moreover, a variable density profile was admitted as compared to the incompressible Schwarzschild sphere.

\end{document}